%% file: main.tex
\documentclass{article}

\usepackage{microtype}
\usepackage{graphicx}
\usepackage{subcaption}
\usepackage{booktabs} 

\usepackage{hyperref}

\usepackage[preprint]{icml2026}

\usepackage{amsmath}
\usepackage{amssymb}
\usepackage{mathtools}
\usepackage{amsthm}

\usepackage[capitalize,noabbrev]{cleveref}

\theoremstyle{plain}

\theoremstyle{definition}

\theoremstyle{remark}

\usepackage[textsize=tiny]{todonotes}

\input{preamble}

\begin{document}

\twocolumn[
  \icmltitle{Agentic Harness for Real-World Compilers}



  \icmlsetsymbol{equal}{*}

  \begin{icmlauthorlist}
    \icmlauthor{Yingwei Zheng}{equal,sustech}
    \icmlauthor{Cong Li}{equal,ethz}
    \icmlauthor{Shaohua Li}{cuhk}
    \icmlauthor{Yuqun Zhang}{sustech}
    \icmlauthor{Zhendong Su}{ethz}
  \end{icmlauthorlist}

  \icmlaffiliation{sustech}{Southern University of Science and Technology}
  \icmlaffiliation{ethz}{ETH Zurich}
  \icmlaffiliation{cuhk}{The Chinese University of Hong Kong}

  \icmlcorrespondingauthor{Shaohua Li}{shaohuali@cuhk.edu.hk}
  \icmlcorrespondingauthor{Yuqun Zhang}{zhangyq@sustech.edu.cn}

  \icmlkeywords{Machine Learning, ICML}

  \vskip 0.3in
]



\printAffiliationsAndNotice{\icmlEqualContribution}

\input{sections/expdata}

\begin{abstract}
\input{sections/abstract}
\end{abstract}

\input{sections/introduction}
\input{sections/platform}

\input{sections/evaluation}
\input{sections/relatedwork}
\input{sections/conclusion}



\bibliography{main}
\bibliographystyle{icml2026}

\newpage
\appendix
\onecolumn
\input{sections/implementation}
\input{sections/more_evaluation}
\input{sections/more_examples}

\end{document}

%% file: preamble.tex
\usepackage[utf8]{inputenc}
\usepackage[T1]{fontenc}

\usepackage{caption}
\usepackage{inconsolata}
\usepackage{amssymb,amsmath,amsfonts}
\usepackage{soul}
\usepackage{xspace}
\usepackage{relsize}
\usepackage{listings}
\usepackage{enumitem}
\usepackage{sepnum}
\usepackage{multirow}
\usepackage{multicol}
\usepackage{makecell}
\usepackage{colortbl}

\setlist[itemize]{leftmargin=1.5em}
\setlist[enumerate]{leftmargin=1.5em}

\definecolor{ColorLstHighlight}{RGB}{209,214,250}
\definecolor{ColorCodeBackground}{RGB}{250,250,250}
\definecolor{ColorCodeKeyword}{RGB}{0,0,255}
\definecolor{ColorCodeComment}{rgb}{0,0.6,0}
\definecolor{ColorCodeString}{rgb}{0.58,0,0.82}
\definecolor{ColorCodeNumber}{rgb}{0.5,0.5,0.5}
\definecolor{ColorAlgoKeyword}{RGB}{0,0,255}
\definecolor{ColorAlgoComment}{rgb}{0,0.6,0}
\definecolor{ColorAlgoNumber}{rgb}{0.5,0.5,0.5}
\definecolor{ColorDiffAddition}{rgb}{0,0.6,0}
\definecolor{ColorDiffDeletion}{RGB}{255,25,36}

\newcommand{\codefontsize}{\defaultcodefontsize}
\lstdefinestyle{defaultlststyle}{
  backgroundcolor=\color{ColorCodeBackground},
  basicstyle=\ttfamily\codefontsize,
  keywordstyle=\color{ColorCodeKeyword}\bf\ttfamily,
  commentstyle=\color{ColorCodeComment}\bf\ttfamily,
  numberstyle=\tiny\color{ColorCodeNumber},
  stringstyle=\color{ColorCodeString},
  lineskip=-1pt,
  breakatwhitespace=false,
  columns=fullflexible,
  keepspaces=true,
  showspaces=false,
  showstringspaces=false,
  breaklines=true,
  breakatwhitespace=true,
  captionpos=b,
  numbers=left,
  numbersep=4pt,
  xleftmargin=0pt,
  frame=tb,
  framerule=0.5pt,
  basewidth=0.5em,
  escapechar=§,
  escapeinside={(*@}{@*)},
  showtabs=false,
  tabsize=2,
  language=C,
  morekeywords={find,grep,read,list,reset,preview,docs,langref,code,eval,debug,edit,test,str,write,bash,todo,build,interp,opt,exec,comp,veri,veri2}
}
\newcommand{\code}[1]{\texttt{\smaller[0.0]{#1}}}
\newcommand{\lstcode}[1]{\lstinline[basicstyle=\ttfamily\small]|#1|}

\usepackage{lstlinebgrd}
\usepackage{pgffor}

\newcommand{\highlightmultilines}[1]{%
  \lstset{linebackgroundcolor={%
    \xdef\lstHL{\noexpand\color{ColorCodeBackground}}%
    \foreach \x in {#1}{%
      \ifnum\value{lstnumber}=\x%
        \xdef\lstHL{\noexpand\color{ColorDiffAddition!12!white}}%
      \fi%
      \ifnum\value{lstnumber}=-\x%
        \xdef\lstHL{\noexpand\color{ColorDiffDeletion!12!white}}%
      \fi%
    }%
    \lstHL}}}

\lstnewenvironment{lstcodediff}[1][]{
  \lstset{#1}
}{
}

\lstdefinestyle{promptlststyle}{
  basicstyle=\ttfamily\codefontsize,
  breakatwhitespace=true,
  breakindent=0pt,
  lineskip=-1pt,
  numbers=none,
  numbersep=0pt,
  xleftmargin=6pt,
  framexleftmargin=1em,
  language={}
}

\lstnewenvironment{llmprompt}{
  \lstset{style=promptlststyle}
}{
  \lstset{style=defaultlststyle}
}

\newcommand{\ourplatform}[1][]{\code{llvm-harness#1}\xspace}
\newcommand{\ourbenchmark}[1][]{\code{llvm-bench#1}\xspace}
\newcommand{\ouragent}[1][]{\code{llvm-autofix-mini#1}\xspace}
\newcommand{\mswe}[1][]{\code{mini-SWE-agent#1}\xspace}
\newcommand{\swev}[1][]{\code{SWEV#1}\xspace}

\newcommand{\checkpass}{\color{green!60!black} $\checkmark$}
\newcommand{\checkfail}{\color{red!70!black} $\times$}

\newcommand{\smalltitle}[2][.]{\smallskip\noindent\textbf{#2#1} }
\newcommand{\smallsmalltitle}[2][.]{\smallskip\noindent\emph{#2#1} }
\newcommand{\numwithcomma}[1]{\sepnum{.}{,}{}{#1}}


%% file: sections/expdata.tex
\newcommand\numberofissues[0]{334\xspace}
\newcommand\numberofcrashes[0]{222\xspace}
\newcommand\crashesratio[0]{66.5\%\xspace}
\newcommand\numberofmiscompilations[0]{112\xspace}
\newcommand\miscompilationsratio[0]{33.5\%\xspace}
\newcommand\numberofissueslive[0]{229\xspace}
\newcommand\numberofcrasheslive[0]{160\xspace}
\newcommand\crashesratiolive[0]{69.9\%\xspace}
\newcommand\numberofmiscompilationslive[0]{69\xspace}
\newcommand\miscompilationsratiolive[0]{30.1\%\xspace}
\newcommand\tablecomponentdistrib[0]{
LoopVectorize & 79 (23.7\%) \\
SLPVectorizer & 78 (23.4\%) \\
InstCombine & 53 (15.9\%) \\
ScalarEvolution & 15 (4.5\%) \\
VectorCombine & 13 (3.9\%) \\
ValueTracking & 7 (2.1\%) \\
IR & 6 (1.8\%) \\
ConstraintElimination & 6 (1.8\%) \\
LoopPeel & 4 (1.2\%) \\
InstructionSimplify & 4 (1.2\%) \\
Others & 69 (20.7\%) \\

}
\newcommand\tableexpresult[0]{
GPT 5 & 21.0 & 61.4 & 32.7 & {\bf 51.5 } & {\bf 76.3 } & {\bf 42.4 } \\
GPT 4o & 8.3 & 59.2 & 29.9 & 12.2 & 42.5 & 21.8 \\
DeepSeek V3.2 & 38.9 & 64.5 & 36.2 & 10.5 & 19.6 & 13.7 \\
Qwen3 Max & 24.4 & 64.2 & 36.0 & 35.8 & 64.4 & 35.6 \\
Gemini 2.5 Pro & 9.2 & 55.0 & 23.8 & 14.4 & 48.6 & 23.6 \\

}
\newcommand\multifilefixcount[0]{35\xspace}
\newcommand\multifilefixratio[0]{10.5\%\xspace}
\newcommand\multifuncsinglefilefixcount[0]{44\xspace}
\newcommand\multifuncsinglefilefixratio[0]{13.2\%\xspace}
\newcommand\singlefilefixcount[0]{299\xspace}
\newcommand\singlefilefixratio[0]{89.5\%\xspace}
\newcommand\singlefuncfixcount[0]{255\xspace}
\newcommand\singlefuncfixratio[0]{76.3\%\xspace}
\newcommand\singlehunkfixcount[0]{191\xspace}
\newcommand\singlehunkfixratio[0]{57.2\%\xspace}
\newcommand\componentscount[0]{184\xspace}
\newcommand\llvmfiles[0]{\numwithcomma{937}\xspace}
\newcommand\llvmlocs[0]{\numwithcomma{676308}\xspace}
\newcommand\relatedcomponentscount[0]{64\xspace}
\newcommand\relatedcomponentscountlive[0]{43\xspace}
\newcommand\relatedsubcomponentscount[0]{176\xspace}
\newcommand\relatedsubcomponentscountlive[0]{175\xspace}
\newcommand\goldeneditedfiles[0]{1.17\xspace}
\newcommand\goldeneditedfuncs[0]{1.62\xspace}
\newcommand\goldeneditedlines[0]{17.06\xspace}

%% file: sections/abstract.tex
Compilers are critical to modern computing, yet fixing compiler bugs is difficult. While recent large language model (LLM) advancements enable automated bug repair, compiler bugs pose unique challenges due to their complexity, deep cross-domain expertise requirements, and sparse, non-descriptive bug reports, necessitating compiler-specific harnesses. To bridge the gap, we introduce \ourplatform, the first harness designed to assist LLM agents in understanding and fixing compiler bugs. Our current focus is on the middle end of LLVM, one of the most widely used compiler infrastructures. Central to \ourplatform are agent-friendly LLVM tools, a benchmark \ourbenchmark of \numberofissues reproducible LLVM middle-end bugs, and a tailored mini agent \ouragent for fixing LLVM middle-end bugs automatically. We evaluate five frontier models and find that they exhibit a performance decline when tackling compiler bugs with the state-of-the-art agent. With \ourplatform' enhancement, their performance improves by 62\%. Our specialized mini agent \ouragent further outperforms the \ourplatform-enhanced state-of-the-art by 22\%. This emphasizes the necessity for specialized harnesses like ours to assist LLMs in compiler engineering tasks. Despite promising results, our expert review also reveals several open challenges that remain when applying LLMs for compiler engineering tasks.
GitHub: \url{https://github.com/dtcxzyw/llvm-harness}

%% file: sections/introduction.tex

\section{Introduction}\label{sec:introduction}

Compilers serve as the foundational infrastructure for modern computing.
They translate source code into efficient machine instructions.
Virtually every piece of software,
ranging from low-level OS kernels, machine learning (ML) frameworks, to high-level web applications and ML models,
relies on them for correct compilation and performance optimization.
Hence, it is critical to repair and maintain compilers as soon as any issues arise.
Recent advances in large language models (LLMs) present great potential in this direction~\cite{sweagent,crashfixer}.

\begin{figure*}[tb]
    \renewcommand{\codefontsize}{\tiny}
    \centering
    \begin{subfigure}[t]{0.30\textwidth}
        \input{figures/code_django_11848}
        \caption{Django \#11848}
        \label{fig:django-issue-example}
    \end{subfigure}
    \hfill
    \begin{subfigure}[t]{0.35\textwidth}
        \input{figures/code_llvm_99899}
        \caption{LLVM \#99899}
        \label{fig:llvm-crash-example}
    \end{subfigure}
    \hfill
    \begin{subfigure}[t]{0.32\textwidth}
        \input{figures/code_llvm_152824}
        \caption{LLVM \#152824}
        \label{fig:llvm-miscomp-example}
    \end{subfigure}
    \caption{
        \textbf{Compiler issues, lacking descriptive information, are challenging to diagnose and repair}.
        This compares two LLVM issues (crash and miscompilation) and a common software issue from Django;
        all issues are simplified for brevity.
    }
    \label{fig:issue-comparison-example}
\end{figure*}

However, unlike common software, compilers represent a unique category of large-scale, complex software systems.
Their internal complexity, marked by sophisticated intermediate representation (IR) and numerous optimizations, renders them exceptionally challenging to maintain and repair.
Common software bugs, such as \Cref{fig:django-issue-example}, typically include natural language descriptions to aid developers' understanding,
but such descriptive information is absent in compiler bugs, complicating diagnosis and repair:
\begin{itemize}
    \item Compiler \emph{crash} bugs (\Cref{fig:llvm-crash-example}), come with a test case \code{@test()} (known as a \emph{reproducer} in compiler engineering) and a stack trace that pinpoints where the compiler crashes while compiling the reproducer.

    \item Compiler \emph{miscompilation} bugs (\Cref{fig:llvm-miscomp-example}), offer a reproducer \code{@src()} and occasionally an input-output pair (known as a \emph{counterexample}) that illustrates how the compiler incorrectly compiles the reproducer into faulty code \code{@tgt()}, leading to wrong output given the input.

    \item Other compiler bugs are even more intricate, typically presenting only a reproducer:
    \emph{slow compilations} and \emph{compiler hangs} indicate compilers unexpectedly take a long time or forever to compile a reproducer;
    \emph{missed optimizations} denote compilers failing to apply certain beneficial code transformations on a reproducer.
\end{itemize}
Furthermore, understanding compilers requires specialized tooling and expertise rarely used in general software development, including lexing/parsing, type systems, IR design/optimization, and code generation, which typically take years for human engineers to master.
While existing platforms such as SWE-bench~\cite{swebench} and SWE-agent~\cite{sweagent} effectively connect LLMs to standard bash tools for general software engineering tasks, they exhibit limited efficacy in compiler engineering.

To explore the capabilities of LLMs in the context of resolving compiler issues, we put our focus on LLVM, one of the most widely used compilers powering languages from traditional C/C++/Rust to modern Triton/Mojo.
At present, we target its middle end, which transforms an LLVM IR program into an optimized version via various optimization components.
We prioritize this target due to the expressive, well-defined LLVM IR~\cite{llvmlangref} and the complexity of its numerous, target-independent, and error-prone optimizations~\cite{llvmpasses,understandingcompilers}.

\smalltitle{\ourplatform}
We introduce \ourplatform, the first specialized harness designed to assist autonomous agents in understanding and fixing compiler bugs.
\ourplatform is divided into three parts (\Cref{sec:platform}):

\begin{itemize}
    \item A set of LLVM-specific tools for building, reproducing, exploring, debugging, and testing LLVM.
    These tools provide an agent-friendly interface, eliminating unnecessary technical details and allowing agents to focus on the core aspects of bug localization and repair.

    \item A benchmark called \ourbenchmark consisting of \numberofissues reproducible LLVM middle-end bugs, each accompanied by around 1.4 reproducers and 722 regression tests.
    They are separated into three splits based on difficulty: \code{easy} (76.3\%), \code{medium} (13.2\%), and \code{hard} (10.5\%).
    The benchmark focuses on the two most common types of compiler bugs: crashes and miscompilations,
    as performance issues are often less prioritized by LLVM developers.
    Additionally, we maintain \ourbenchmark[ live], a continuously updated \ourbenchmark subset including only issues from the past year.

    \item A mini agent called \ouragent tailored for fixing LLVM middle-end bugs.
    Unlike general, state-of-the-art auto-fixing agents such as \mswe~\cite{minisweagent,sweagent}, ours draws on our real-world experience and domain knowledge in repairing and maintaining LLVM.
    We organized it into four stages: ``Setup $\to$ Reason $\to$ Generate $\to$ Validate'', where the reasoning stage and the generation stage construct their agentic loops leveraging LLVM-specific tools from \ourplatform.
    \ouragent serves as a demonstration of the harness and as a more promising baseline for future LLVM-specific auto-fixing agents.
\end{itemize}

Leveraging \ourplatform, we conduct a systematic evaluation (\Cref{sec:eval-results,sec:rq4}) of five frontier models to explore and understand their ability to resolve LLVM middle-end issues in \ourbenchmark[ live].
The evaluated models include GPT 5, Gemini 2.5 Pro, DeepSeek V3.2, and Qwen 3 Max.

\smalltitle{Basic findings from benchmarking experiment}
We integrate the state-of-the-art agent \mswe, which is designed to bridge the gap between LLMs and common software engineering tasks, with additional \ourplatform tools to establish a baseline agent.
In our experiments with \mswe, we find that frontier models struggle with addressing LLVM middle-end bugs in \ourplatform, despite their demonstrated success with general software bugs in SWE-bench Verified.
Specifically, we observe a decline of over 35\% (averaging 60\%) in resolution rates when replacing SWE-bench Verified with \ourbenchmark[ live].
The best-performing model, DeepSeek V3.2, can resolve only 38\% of issues--43\%, 32\%, and 17\% of \code{easy}, \code{medium}, and \code{hard}, respectively--compared to its 60\% score in SWE-bench Verified~\cite{swebenchleaderboard}.

We also find that \ourplatform offers critical support.
On the one hand, removing \ourplatform tools from \mswe results in an additional 38\% decrease in resolution rate for its best-performing model, DeepSeek V3.2.
On the other hand, our tailored agent \ouragent outperforms the enhanced \mswe by 22\% per model.
With our agent, the best-performing model is GPT 5, which resolves 52\% of issues, with 59\%, 35\%, and 21\% of the \code{easy}, \code{medium}, and \code{hard} split, respectively.
This highlights the usefulness of \ourplatform and the necessity of developing compiler-specific harnesses and agents.

\smalltitle{More findings from expert review study}
To better understand the \emph{true} capabilities of frontier models, we recruit active LLVM developers and maintainers to review all agent-generated patches.
Specifically, experts assess whether each agent-generated, benchmark-passing patch is a valid fix for the corresponding bug.
The findings are twofold.

First, \ouragent produces 38\% more valid patches than \mswe, with \ouragent's best-performing model, GPT 5, outperforming \mswe's top model DeepSeek V3.2 by 40\%.
Additionally, the percentage of valid patches among benchmark-passing patches is 24\% higher for \ouragent compared to \mswe, further highlighting the utility of \ourplatform.

Second, we find that \emph{truly resolving} LLVM middle-end issues is exceptionally difficult:
the true capability of frontier models with both agents is consistently below 21\%.
In the \code{hard} split, only \ouragent (with GPT 5) manages to resolve a single issue.
To investigate why the remaining patches are invalid, we ask experts to document the cause of each invalid patch when reviewing.
This results in several actionable directions for future work (\Cref{sec:llm-for-compilers}).

\smalltitle{Contributions}
Our primary contribution is the specialized harness \ourplatform and the accompanying \ourbenchmark benchmark,
which are pivotal for enabling various agents to work with LLVM effectively.
\ouragent is designed as a demonstration to show \ourplatform' necessity and utility, and to
serve as a baseline for future LLVM auto-fixing agents.
As a secondary contribution, we conduct a systematic evaluation of frontier models,
which uncovers insightful findings and open challenges for future work to address when adopting LLMs to resolve LLVM and compiler bugs.
We plan to publicly release \ourplatform upon the paper's publication.
They are currently included in the supplementary material.

%% file: figures/code_django_11848.tex


     


\begin{llmprompt}
django.utils.http.parse_http_date two digit year check is incorrect

Description:

(last modified by Ad Timmering)
     
RFC 850 does not mention this, but in RFC 7231 (and there's something similar in RFC 2822), there's the following quote: Recipients of a timestamp value in ... 

Current logic is hard coded to consider 0-69 to be in 2000-2069, and 70-99 to be 1970-1999, instead of comparing versus the current year.
\end{llmprompt}

%% file: figures/code_llvm_99899.tex
\begin{lstlisting}[language=llvm,numbers=none]
;; opt -mcpu=z16 -O3 ...
define i1 @test(i64 %0, i64 %1, ptr %2) {
  %a = getelementptr i8, ptr null, i64 %0
  %b = getelementptr i8, ptr null, i64 %1
  %4 = icmp ult ptr %a, %b
  ...
  %x = icmp ult ptr %u, %2
  %r = and i1 %y, %x
  ret i1 %r
}
;; Assertion: isIntOrIntVectorTy() failed.
;;   Original type expected to be a vector
;;   of integers or a scalar integer. 
;; Stack trace (#14--#1 omitted):
;;   #15 slpvectorizer::BoUpSLP::getEntryCost
;;   #14 ...
\end{lstlisting}

%% file: figures/code_llvm_152824.tex

\begin{lstlisting}[language=llvm,numbers=none]
;; opt -passes=instsimplify -S ...
define half @src(half noundef %c) {
  %a = call half @llvm.fabs.f16(half %c)
  ...
  %b = call half @llvm.fabs.f16(half %d)
  ret half %b
} ;; Incorrectly optimized into =>
define half @tgt(half noundef %c) {
  %a = call half @llvm.fabs.f16(half %c)
  ...
  %d = select i1 %c, half %c, half 0.0
  ret half %d
}
;; Counterexample input: %x = #x8001
;;   Source @src output: #x0001
;;   Target @tgt output: #x8001
\end{lstlisting}

%% file: sections/platform.tex
\section{The \ourplatform Harness}\label{sec:platform}

\ourplatform is an off-the-shelf agent harness designed to assist auto-fixing agents in understanding and repairing LLVM middle-end bugs, with the goal of covering all types of LLVM bugs.
The task is to generate a patch that resolves the bug and passes all relevant LLVM IR tests, using one or more reproducers that faithfully replicate the bug.
In this section, we elaborate on the implementation details of
the harness tooling, the benchmark, and the mini agent.

\subsection{The Harness Tooling}\label{sec:tooling}

We wrap common LLVM tasks into agent-accessible tools so that agents can focus on bug repair while the harness handles environment control and validation.
Tools belong to the following categories;
detailed descriptions are provided in \Cref{sec:enabled-tools-in-eval,sec:other-harness-tools} of our supplementary material's \code{appendix.pdf}.

\smalltitle{Setup \& Build}
The harness handles build configurations and provides tools for building LLVM with a specific commit and/or target, for example, building for \code{x86} with debug information and assertion checks.
The agent can therefore avoid the task of building large systems, a common challenge for autonomous agents~\cite{compileagent}.

\smalltitle{Reproduce \& Cause}
The harness is capable of validating whether a given bug can be triggered faithfully and providing direct causes of reproducible bugs.
It sets up LLVM with the bug-containing version, builds it, and runs the reproducer with \code{opt}, i.e., the LLVM middle-end optimizer.
For crash bugs like \Cref{fig:llvm-crash-example}, it checks whether the crash occurs and provides the stack trace to agents, with uninteresting frames eliminated.
For miscompilations like \Cref{fig:llvm-miscomp-example}, it ensures \code{opt} exits normally and validates LLVM's optimization by applying \code{alive-tv}~\cite{alive,alive2}, i.e., LLVM's middle-end translation validator;
this will output a counterexample demonstrating the bug if successful.

\smalltitle{Explore \& Debug}
The harness offers search tools for exploring LLVM's static and dynamic internals.
Static information includes commit-specific code, documentation, and LLVM IR's specification~\cite{llvmlangref} through \code{grep}- and \code{find}-like tools.
Dynamic information enables pausing LLVM and inspecting its intermediate states, for example, reading variables and evaluating expressions at a breakpoint, with respect to a reproducer through the \code{gdb} debugger.

\smalltitle{Edit \& Patch}
With \ourplatform, agents can edit LLVM, preview and revert edits, and submit edits to the harness.
When necessary, \ourplatform checks agents' edits to prevent irrelevant code changes, such as changes in the frontend, backend, or other tools under the LLVM umbrella.

\smalltitle{Test \& Validate}
The correctness of the submitted patch can be validated, and feedback can be provided.
\ourplatform will rebuild LLVM and execute relevant tests, typically including reproducers themselves and regression tests of the associated middle-end component such as \code{InstCombine}.
Extensive testing involving regression tests from other components such as \code{SLPVectorizer}
and differential testing of the IR post-optimization (if the golden patch is accessible) are also available.
This validation is supported by various compiler-specific engineering tools, such as \code{opt}, \code{alive-tv}, and \code{llvm-lit} (or \code{FileCheck}).
We do not include random testing tools such as \code{csmith}~\cite{csmith} and \code{reify}~\cite{reify} because they struggle to detect bugs within merely a few hours.
Submitted patches can be validated online (i.e., during agent execution) or offline (i.e., after agent execution).

\smalltitle{Benchmark \& Evaluate}
Based on the above tooling, \ourplatform provides scripts to automatically benchmark and evaluate an agent with our benchmark \ourbenchmark.
These scripts output the validated patches (or errors for failed runs), cost and overhead, and trajectories.


\subsection{The \ourbenchmark Benchmark}\label{sec:benchmark}

We build a benchmark to explore and understand the capabilities of frontier models or auto-fixing agents.

\smalltitle{Construction}
We construct \ourbenchmark \emph{automatically} by leveraging the \ourplatform tooling.
The construction process consists of three stages:
(1) issue collection, (2) reproducer validation, and (3) golden patch validation.

\smallsmalltitle{Stage I: Issue collection}
We collect LLVM's fixed issues on GitHub and the corresponding commits.
For each issue, we include the following details.

\begin{itemize}
    \item \emph{Type}:
    Either miscompilation or crash in the middle-end, indicated by the developer-annotated labels.
    Issues lacking relevant labels or bearing labels such as \code{wontfix}, \code{duplicate}, and \code{invalid} are excluded.
    
    \item \emph{Fixing Commit}:
    The commit that resolves the issue.
    In LLVM, valid middle-end fixing commits usually alter code in \code{llvm/lib/} or \code{llvm/include/},
    and usually introduce new tests in \code{llvm/test/} or modify existing ones.
    Issues are excluded if no such commit exists, is deemed invalid, or reverts LLVM to a previous commit.
    
    \item \emph{Reproducers}:
    One or more LLVM IR files and associated \code{opt} command.
    These LLVM IR programs are extracted using \code{llvm-extract} from the fixing commit, based on code changes relevant to tests in \code{llvm/test/}.
    
    \item \emph{Golden Patch}:
    Code changes in \code{llvm/lib/} or \code{llvm/include/}, as provided in the fixing commit.
    Stage III will validate the correctness of this patch.

    \item \emph{Base Commit}:
    The commit able to reproducing the issue.
    This is the parent commit of the fixing commit.
    If the parent commit is not buildable, we trace \code{git} history of \code{main} branch to identify the most recent buildable commit;
    we will update the golden patch and reproducers accordingly.
    This commit will be validated in Stage II.
    
    \item \emph{Metadata}:
    This includes the issue's title, description, timestamps, additional labels, and developer discussions.
    Although this information should not be utilized by auto-fixing agents since it may provide hints on bug fixes, it is invaluable for understanding the issue.
\end{itemize}
Note that when multiple issues are closed by the same commit, only the oldest issue is retained,
because recent issues typically refer to cases of duplication that lack appropriate \code{duplicate} labels due to LLVM developers' oversight.

\smallsmalltitle{Stage II: Reproducer validation}
We validate whether the extracted reproducers can be reproduced on the base commit leveraging tools for \textbf{Reproduce \& Cause} (\Cref{sec:tooling}).
Issues that cannot be reproduced are excluded.

\smallsmalltitle{Stage III: Golden patch validation}
We validate whether the golden patch is valid by executing \ourplatform tools for \textbf{Test \& Validate} (\Cref{sec:tooling}) on the base commit.
Issues with the golden patch that cannot be validated are excluded.

\begin{table}[tb]
    \centering
    \scriptsize
    \setlength{\tabcolsep}{.35em}
    \caption{
        \textbf{\ourbenchmark includes 334 reproducible issues}.
        ``\#C/\#M'' is the number of crashes/miscompilations; ``\#Comp'' the number of affected components.
        ``\#Rep'' and ``\#Reg'' are the average number of reproducers and component-specific regression tests per issue.
        ``Fixes'' report the average edited lines, functions, and files per golden patch.
    }
    \label{tab:dataset-stats}
    \begin{tabular}{ccrcccrcc}
        \toprule
        & & & & \multicolumn{2}{c}{\bf Tests} & \multicolumn{3}{c}{\bf Fixes} \\ 
        \cmidrule(lr){5-6}
        \cmidrule(lr){7-9}
        \bf Name & \bf Split & \bf \#Bugs (\#C/\#M) & \bf \#Comp & \it \#Rep & \it \#Reg & \it \#Lines & \it \#Funcs & \it \#Files \\
        \midrule
\multirow{4}{*}{\code{full}} & \code{full} & 334 (222/112) & 64 & \numwithcomma{1.4} & \numwithcomma{722.2} & \numwithcomma{17.1} & \numwithcomma{1.6} & \numwithcomma{1.2} \\
\cmidrule(lr){2-9}
 & \code{easy} & 255 (176/79) & 48 & \numwithcomma{1.4} & \numwithcomma{734.1} & \numwithcomma{9.1} & \numwithcomma{1.0} & \numwithcomma{1.0} \\
 & \code{medium} & 44 (26/18) & 18 & \numwithcomma{1.4} & \numwithcomma{626.5} & \numwithcomma{38.8} & \numwithcomma{2.6} & \numwithcomma{1.0} \\
 & \code{hard} & 35 (20/15) & 25 & \numwithcomma{1.7} & \numwithcomma{756.6} & \numwithcomma{47.6} & \numwithcomma{4.9} & \numwithcomma{2.7} \\
\midrule
\multirow{4}{*}{\code{live}} & \code{full} & 229 (160/69) & 43 & \numwithcomma{1.5} & \numwithcomma{743.7} & \numwithcomma{18.5} & \numwithcomma{1.7} & \numwithcomma{1.2} \\
\cmidrule(lr){2-9}
 & \code{easy} & 172 (124/48) & 31 & \numwithcomma{1.5} & \numwithcomma{769.8} & \numwithcomma{9.0} & \numwithcomma{1.0} & \numwithcomma{1.0} \\
 & \code{medium} & 34 (21/13) & 13 & \numwithcomma{1.4} & \numwithcomma{646.7} & \numwithcomma{41.6} & \numwithcomma{2.6} & \numwithcomma{1.0} \\
 & \code{hard} & 23 (15/8) & 20 & \numwithcomma{1.8} & \numwithcomma{691.7} & \numwithcomma{55.7} & \numwithcomma{5.1} & \numwithcomma{2.8} \\
        \bottomrule
    \end{tabular}
\end{table}

\smalltitle{Statistics}
As of August 26, 2025 when we started benchmark construction, we successfully collected \numberofissues issues starting from January 1, 2024.%
\footnote{
We did not consider issues prior to this date as old issues might be included in the recent LLMs' training data.
Moreover, \ourplatform[] is continually updated:
When we submitted this paper, we had extended it to 474 reproducible issues.
}
This includes \numberofcrashes crashes and \numberofmiscompilations miscompilations.
An example crash issue is presented in \Cref{sec:example-issue} in our supplementary material's appendix.
The upper half of \Cref{tab:dataset-stats} with ``Name: \code{full}'' displays the statistics.

\begin{figure}[tb]
    \centering
    \includegraphics[width=\linewidth]{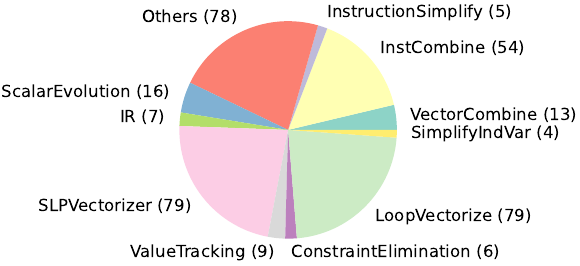}
    \caption{Distribution of Affected Components in \ourbenchmark}
    \label{fig:component-distribution}
\end{figure}

\smallsmalltitle{Affected components}
These issues directly affect \relatedcomponentscount out of \componentscount%
\footnote{Counted based on LLVM 21.1.0's \code{-O3} optimization pipeline.}
middle-end components in LLVM, with \relatedsubcomponentscount components affected indirectly.
\code{LoopVectorize}, \code{SLPVectorizer}, and \code{InstCombine}--three components known to be the most erroneous~\cite{csmith,yarpgen1,yarpgen2,creal}--are affected by the greatest number of issues.
\Cref{fig:component-distribution} presents a breakdown; the full list is shown in \Cref{sec:full-list-affected-componnets} of our supplementary material's appendix.
These components span \llvmfiles C++ files and \numwithcomma{\llvmlocs} lines of LLVM code.
The benchmark lacks issues from other components, as these issues predate January 1, 2024.

\smallsmalltitle{Reproducers \& tests}
On average, each issue is coupled with 1.41 reproducers, \numwithcomma{722} component-specific regression tests, and more than \numwithcomma{10000} regression tests from other components.
Reproducers and regression tests are minimal LLVM IR programs reduced by automatic tools and LLVM developers.
Each reproducer has 49.74 LoC on average.

\smallsmalltitle{Fixes \& patches}
Resolving these LLVM issues is challenging.
When LLVM developers initially submitted these patches for code review, fewer than 50\% were accepted without further bug fixes;
the remainder required corrections based on reviewers' feedback and subsequent resubmission.
In the end, every developer-provided golden patch involves around \goldeneditedlines edited lines, \goldeneditedfuncs functions, and \goldeneditedfiles files.

\smalltitle{Three splits}
Based on the difficulty in resolving these issues, we divide \ourbenchmark into three splits:
\code{easy} issues (\singlefuncfixcount, \singlefuncfixratio), which can be resolved by updating a single function;
\code{medium} issues (\multifuncsinglefilefixcount, \multifuncsinglefilefixratio), which require changes to multiple functions within the same file; and
\code{hard} issues (\multifilefixcount, \multifilefixratio), which necessitate changes across multiple files.
As the splits become more difficult, code fixes become more complex.
For example, the \code{hard} split involves edits to around five functions spanning around three files on average.

\smalltitle{One subset: \ourbenchmark[ live]}
Evaluating agents across the entire \ourbenchmark can be both time-consuming and costly;
our experiments indicate that even the best-performing model requires at least 15 minutes and 1.3 million tokens to fix an issue.
On the other hand, old issues may become outdated or less useful for the most recent LLMs.
Therefore, we create a \emph{continually} updated \code{live} subset that always includes \ourbenchmark issues from \emph{the latest year}.
At present, as shown in the lower half of \Cref{tab:dataset-stats}, it has \numberofissueslive issues starting from August 26, 2024, spanning \relatedcomponentscountlive directly affected components and \relatedsubcomponentscountlive indirectly affected components.
Given the evolving nature of compilers, such as LLVM, an agent's performance on \ourbenchmark[ live] can reflect its compatibility and effectiveness with the latest compiler versions.
Therefore, this subset aims to mitigate the data leakage problem, a common challenge for benchmarking LLMs~\cite{benchmarkprob,openaigiveupswe}.

\subsection{The \ouragent Agent}\label{sec:agent}

To demonstrate the practicality and benefits of our harness, we design a mini agent that takes advantage of the tooling.
Unlike typical auto-fixing agents~\cite{sweagent,crashfixer} that resolve software bugs based solely on static information from bug reports and software repositories, \ouragent additionally utilizes LLVM's runtime information obtained through the reproducer.
We structure it into a four-stage agent:
``Setup $\to$ Reason $\to$ Generate $\to$ Validate''.
The first two stages perform root cause analysis to explore information beneficial for the third stage;
the final stage handles post-generation, offline patch validation.
If the agent unexpectedly exits during the first three stages (e.g., due to reaching the token limit) or if the patch submitted to the final stage is invalid, it results in a failure.
Otherwise, a validated patch, which we call \emph{a passed patch}, is output.

\smallsmalltitle{Stage I: Setup}
\ouragent configures the environment using \ourplatform by
validating the reproducibility of the reproducer,
launching LLVM under \code{gdb} with the reproducer, and
pausing it at specific breakpoints.
For crash bugs, it pauses before the crashing function; for miscompilation bugs, before the first transformation.
These breakpoints are chosen because, based on our experience in fixing LLVM bugs, minimal reproducers usually indicate that the root cause is in the neighborhood of these points.
This pause allows our agent to examine LLVM's internal state and detect errors through dynamic debugging.
This stage also infers the erroneous component, which is then passed to the LLM in the next \emph{Reason} stage and used for component-specific online testing.

\smallsmalltitle{Stage II: Reason}
Our agent enters a ReAct~\cite{react} agentic loop to deduce the root cause.
Within this loop, it has access to exploration-only tools.
It may invoke tools to
jump to other frames such as \lstcode{debug(cmd=frame 3)},
inspect local states such as \lstcode{eval(expr=WidePhi)}, or 
review code or docs such as \lstcode{code(func=llvm::VPTransformState::get)} and \lstcode{docs(func=llvm::VPTransformState::get)}.
The loop concludes once the agent successfully identifies the root cause.

\smallsmalltitle{Stage III: Generate}
Based on the root cause determined in the previous stage, it initiates a new ReAct agentic loop for patch synthesis.
This loop augments the agent with editing and testing capabilities.
It may use editing tools to amend or reset LLVM such as \lstcode{edit(file=.../SLPVectorizer,text=..., replace=...)} and \lstcode{reset()}, or testing tools for online testing such as \lstcode{test()}, with the latter offering feedback on failures.
The loop ends when online testing is successful.

\smallsmalltitle{Stage IV: Validate}
This stage consists of offline testing
to assess the correctness of the patch generated in the last stage.
It outputs the patch for successful validation.

%% file: sections/evaluation.tex
\section{Benchmarking Experiment - Setup}\label{sec:expr-setup}

We evaluate frontier LLMs' performance on \ourbenchmark[ live] using \ourplatform.
We avoid \ourbenchmark in order to maintain a fair, up-to-date evaluation.

\smalltitle{Models}
We select four frontier models released in late 2025 when we started our evaluation:
GPT 5, Gemini 2.5 Pro, DeepSeek V3.2, and Qwen 3 Max.
We further select GPT 4o as a baseline model, released about a year prior to the selected frontier models and before the oldest issue in \ourbenchmark[ live].
All models use the latest versions as of October 1, 2025.%
\footnote{
An exception is DeepSeek V3.2, released on December 1, 2025.  
Our initial experiments (late November to early December) used DeepSeek V3.2 Exp, the then-current version.  
A silent update to V3.2 on December 1 caused mixed results, with some data from V3.2 Exp and some from V3.2, so we repeated the experiment using V3.2 only.
}

\smalltitle{Agents}
SWE-bench Verified~\cite{swebenchverified,swebench} (or \swev) and \mswe~\cite{minisweagent,sweagent} are the \emph{de facto} benchmark and agent for measuring frontier models' agentic capability on common software engineering tasks.
As far as we know, all frontier models, including those from OpenAI, Google, DeepSeek, and Alibaba, report agentic results with them.
We therefore adopt \mswe to \ourplatform as the canonical baseline agent and compare it against our agent \ouragent.

\mswe is given access to all bash tools available in Ubuntu (e.g., \code{grep}, \code{sed}) and LLVM (e.g., \code{opt}, \code{lli}, \code{llvm-lit}), following SWE-bench's Bash Only mode.
We further enhance \mswe with reproducers and the following \ourplatform-specific resources:
(1) The bug reproduction process from \ouragent's Stage I (\Cref{sec:agent}), which identifies erroneous components, the root cause, the stack trace for crash bugs, and a counterexample for miscompilation bugs;
(2) The \lstcode{test} tool for building and testing LLVM, along with relevant feedback;
(3) The post-generation validation process from \ouragent's Stage IV (\Cref{sec:agent}).
In contrast, \ouragent has access to a mini subset of \ourplatform tools:
\lstcode{find}, \lstcode{grep}, \lstcode{list},
\lstcode{read}, \lstcode{edit},
\lstcode{code}, \lstcode{docs}, \lstcode{langref},
\lstcode{debug}, \lstcode{eval},
\lstcode{reset}, \lstcode{preview}, and \lstcode{test}.

We do not use larger, production-oriented agents such as Gemini CLI~\cite{geminicli} and Codex~\cite{codex}.
These agents orchestrate many tools and heuristics, which can \emph{mask} the underlying model behavior:
they may fix bugs primarily through extensive scaffolding rather than the model's own problem-solving ability.
However, both \mswe and \ouragent are designed to be \emph{mini}, making them suitable for isolating and comparing the models' inherent agentic ability.
Following LLVM's policy on AI tools~\cite{llvmaiguide}, we plan to integrate these agents into \ourplatform in the future for real-world LLVM repair and maintenance. 
The system prompts for both agents are in Appendix's \Cref{sec:prompts-ours-reason,sec:prompts-ours-generate,sec:prompts-mswe}.

\smalltitle{Parameters}
We use the following settings for both agents:
temperature 0,
context window \numwithcomma{64000} tokens,
default reasoning effort,
chat limit 500 rounds,
token limit 5 million tokens,
edit limit 25 calls,
and online test limit 25 calls.

\smalltitle{Research questions}
We assess whether the agent can produce patches that successfully resolve bugs.
As mentioned earlier, a patch is considered a \emph{passed patch} if it passes all applied online and offline tests, and therefore each resolved issue is associated with one passed patch.
The results are presented as the rate of producing passed patches (``\%Passed = \#PassedPatches / \#Issues'') according to the pass@1 metric.
We then analyze results toward three research questions:

\smallsmalltitle[ (\Cref{sec:rq1}).]{Benchmark and model performance}
Using \mswe, we compare each model's performance on \ourbenchmark[ live] with its performance on \swev.
This helps understand whether \ourbenchmark is more challenging.

\smallsmalltitle[ (\Cref{sec:rq2}).]{Effectiveness of the harness}
To evaluate the effectiveness of \ourplatform, we select the best-performing model from the previous experiment and rerun it with \ourplatform-specific resources excluded.

\smallsmalltitle[ (\Cref{sec:rq3}).]{Baseline comparison and common failures}
We compare the performance of each model with our enhanced \mswe and \ouragent.
This assesses whether \ouragent is a more promising baseline for LLVM middle-end issues.
For unresolved issues, we categorize their common failures.

\section{Benchmarking Experiment - Results}\label{sec:eval-results}

\begin{table}[t]
    \centering
    \scriptsize
    \setlength{\tabcolsep}{.625em}
    \caption{
        \textbf{\ourbenchmark[ live] is a challenging benchmark for current LLMs}.
        Fixing LLVM issues is more difficult than fixing common software issues with \mswe.
        ``\code{SWEV}'' denotes SWE-bench Verified; the data are copied verbatim from the SWE-bench (Bash Only) leaderboard or the respective paper.
        ``\code{full}'', ``\code{easy}'', ``\code{medium}'', and ``\code{hard}'' are splits.
    }
    \label{tab:mswe-results}
    \begin{tabular}{lccccccc}
    \toprule
    & \multicolumn{5}{c}{\bf \%Passed} & \multicolumn{2}{c}{\bf \$Cost} \\
    \cmidrule(lr){2-6}
    \cmidrule(l){7-8}
    \bf Model & \code{SWEV} & \code{full} & \code{easy} & \code{medium} & \code{hard} & \code{SWEV} & \code{full} \\
    \cmidrule(r){1-1}
    \cmidrule(lr){2-2}
    \cmidrule(lr){3-3}
    \cmidrule(lr){4-6}
    \cmidrule(lr){7-7}
    \cmidrule(l){8-8}
GPT 4o & 21.6 & \textcolor{orange!80!black}{\hspace{.5em}8.3}\hspace{.5em}\textcolor{lightgray}{(-61.6{\scriptsize\%})} & \textcolor{red}{\hspace{.5em}9.9} & \textcolor{red!70!black}{\hspace{.5em}5.9} & \textbf{\textcolor{red!70!black}{\hspace{.5em}0.0}} & 1.53 & 1.59 \\
\cmidrule(r){1-1}
\cmidrule(lr){2-2}
\cmidrule(lr){3-3}
\cmidrule(lr){4-6}
\cmidrule(lr){7-7}
\cmidrule(l){8-8}
GPT 5 & 65.0 & \textcolor{orange!80!black}{21.0}\hspace{.5em}\textcolor{lightgray}{(-67.8{\scriptsize\%})} & \textcolor{red}{23.3} & \textcolor{red!70!black}{17.6} & \textbf{\textcolor{red!70!black}{\hspace{.5em}8.7}} & 0.28 & 0.74 \\
Gemini 2.5 Pro & 53.6 & \textcolor{orange!80!black}{\hspace{.5em}9.2}\hspace{.5em}\textcolor{lightgray}{(-82.9{\scriptsize\%})} & \textcolor{red}{10.5} & \textcolor{red!70!black}{\hspace{.5em}5.9} & \textbf{\textcolor{red!70!black}{\hspace{.5em}4.3}} & 0.29 & 0.81 \\
Qwen 3 Max & 69.6 & \textcolor{orange!80!black}{24.4}\hspace{.5em}\textcolor{lightgray}{(-64.9{\scriptsize\%})} & \textcolor{red}{29.1} & \textcolor{red!70!black}{17.6} & \textbf{\textcolor{red!70!black}{\hspace{.5em}0.0}} & -- & 4.32 \\
DeepSeek V3.2 & 60.0 & \textcolor{orange!80!black}{38.9}\hspace{.5em}\textcolor{lightgray}{(-35.2{\scriptsize\%})} & \textcolor{red}{43.0} & \textcolor{red!70!black}{32.4} & \textbf{\textcolor{red!70!black}{17.4}} & 0.03 & 0.08 \\
\cmidrule(r){1-1}
\cmidrule(lr){2-2}
\cmidrule(lr){3-3}
\cmidrule(lr){4-6}
\cmidrule(lr){7-7}
\cmidrule(l){8-8}
\textbf{Mean} & 50.1 & \textcolor{orange!80!black}{17.2}\hspace{.5em}\textcolor{lightgray}{(-60.2{\scriptsize\%})} & \textcolor{red}{19.8} & \textcolor{red!70!black}{12.8} & \textbf{\textcolor{red!70!black}{\hspace{.5em}8.7}} & 0.25 & 0.80  \\
    \bottomrule
    \end{tabular}
\end{table}

\subsection{Benchmark and Model Performance}\label{sec:rq1}
We first evaluate the capabilities of the widely used \mswe in resolving \ourbenchmark[ live] issues and compare its performance with \swev.
In this experiment, we follow established academic and industry practice~\cite{agentless,locagent} to report \swev leaderboard numbers instead of running the benchmark in our own environment.
We use this cross-benchmark comparison to \emph{illustrate} (rather than rigorously quantify) the difficulty of \ourbenchmark and compiler bugs.

\Cref{tab:mswe-results} shows that \mswe faces greater challenges when addressing LLVM middle-end issues in \ourbenchmark compared to common software bugs in \swev.
Even with a budget of five million tokens, which exceeds the typical \mswe settings for \swev, the performance of these models still declines by over 35\% (average: 60.2\%) when \swev is replaced with \ourbenchmark[ live].
Current LLMs' repair capabilities against LLVM issues, although all improved over GPT 4o, are also suboptimal.
The best-performing model is DeepSeek V3.2, yet its performance (89/229) is less than 39\%;
for Gemini 2.5 Pro, only 21 out of 229 (less than 10\%) issues are resolved successfully.
GPT 5 and Qwen 3 Max fall in between (less than 25\%).
These results lag behind their performance on \swev.

\smalltitle{By difficulty}
The situation becomes worse as the benchmark difficulty increases across splits.
Generally, the resolution rate for the \code{medium} and \code{hard} splits is consistently below the average for the \code{full} set, but that for the \code{easy} split is higher.
The baseline GPT 4o model can handle fewer than 9\% of the issues, while for the issues in the \code{hard} split, it is unable to handle any of them.
The four frontier models can partially handle \code{easy} LLVM bugs, underperform in \code{medium} ($\sim$30\% worse), while struggling or even failing in \code{hard} (further $\sim$60\% worse).
It is interesting to note that Qwen 3 Max's performance declines sharply when moving to \code{hard}, whereas the other models degrade more smoothly.

\begin{table}[tb]
    \centering
    \scriptsize
    \setlength{\tabcolsep}{1.125em}
    \caption{
        \textbf{\ourplatform improves \mswe's performance significantly.}
        With \ourplatform, \mswe (with DeepSeek V3.2) achieves a 62\% improvement in \%Passed when addressing LLVM middle-end bugs.
    }
    \label{tab:platform-effectiveness}
    \begin{tabular}{ccccccc}
        \toprule
        & & \multicolumn{3}{c}{\bf Difficulty} & \multicolumn{2}{c}{\bf Bug Type} \\
        \cmidrule(lr){3-5}
        \cmidrule(l){6-7}
        \bf Harness & \code{full} & \code{easy} & \code{medium} & \code{hard} & \code{crash} & \code{miscomp} \\
        \cmidrule(r){1-1}
        \cmidrule(lr){2-2}
        \cmidrule(lr){3-5}
        \cmidrule(l){6-7}
\checkfail & 24.0 & 26.2 & 17.6 & 17.4 & 28.1 & 14.5  \\
\checkpass & 38.9 & 43.0 & 32.4 & 17.4 & 41.9 & 31.9  \\
        \cmidrule(r){1-1}
        \cmidrule(lr){2-2}
        \cmidrule(lr){3-5}
        \cmidrule(l){6-7}
\bf Improv. & \textcolor{green!60!black}{1.6{\scriptsize$\times$}} & \textcolor{green!60!black}{1.6{\scriptsize$\times$}} & \textcolor{green!60!black}{1.8{\scriptsize$\times$}} & \textcolor{green!60!black}{1.0{\scriptsize$\times$}} & \textcolor{green!60!black}{1.5{\scriptsize$\times$}} & \textcolor{green!60!black}{2.2{\scriptsize$\times$}}  \\
        \bottomrule
    \end{tabular}
\end{table}

\subsection{Effectiveness of the Harness}\label{sec:rq2}
To evaluate the effectiveness of \ourplatform, we rerun the previous experiment after excluding \ourplatform.
For this rerun, we use only \mswe with DeepSeek V3.2, as \ouragent is tailored specifically for \ourplatform, and DeepSeek V3.2 demonstrates the best performance in \mswe.
Specifically for \mswe, we remove
(1) the bug reproduction process,
(2) the \lstcode{test} tool, and
(3) the post-validation process,
as detailed in \Cref{sec:expr-setup}.
While \code{alive-tv} and LLVM bash tools such as \code{opt} are included in \ourplatform, we do not consider them to be our contributions; they remain accessible to \mswe.
We compare with the previous experiment to assess its effectiveness.

\Cref{tab:platform-effectiveness} shows that \ourplatform plays a critical role in assisting \mswe to fix LLVM middle-end bugs:
removing it reduces performance by 38\%, bringing \%Passed down to 24.0\%;
conversely, the inclusion of it leads to a significant 62\% improvement in \mswe's performance.
The \code{easy} and \code{medium} splits show performance gains of 64\% and 85\%, respectively.
However, no improvement is observed in the \code{hard} split, likely due to the inherent difficulty of these cases reaching the limit of \mswe in our experimental setup.
The improvement in addressing miscompilations is more pronounced compared to fixing crashes in the middle end.

\begin{table}[t]
    \centering
    \scriptsize
    \setlength{\tabcolsep}{1.3em}
    \caption{
        \textbf{\ouragent is a more promising baseline in repairing LLVM issues than \mswe}.
        Detailed data regarding the cost and overhead is presented in \Cref{tab:cost-overhead}.
    }
    \label{tab:mswe-ours-results}
    \begin{tabular}{lcccc}
    \toprule
    & \multicolumn{2}{c}{\mswe} & \multicolumn{2}{c}{\ouragent} \\
    \cmidrule(lr){2-3}
    \cmidrule(l){4-5}
    \bf Model & \bf \%Passed & \bf \$Cost & \bf \%Passed & \bf \$Cost \\
    \cmidrule(r){1-1}
    \cmidrule(lr){2-2}
    \cmidrule(lr){3-3}
    \cmidrule(lr){4-4}
    \cmidrule(l){5-5}
GPT 4o & \hspace{.5em}8.3 & 1.59 & \textcolor{green!60!black}{12.2}\hspace{.5em}\textcolor{lightgray}{(1.5{\scriptsize$\times$)}} & 2.18  \\
\cmidrule(r){1-1}
\cmidrule(lr){2-2}
\cmidrule(lr){3-3}
\cmidrule(lr){4-4}
\cmidrule(l){5-5}
GPT 5 & 21.0 & 0.74 & \textcolor{green!60!black}{51.5}\hspace{.5em}\textcolor{lightgray}{(2.5{\scriptsize$\times$)}} & 0.59  \\
Gemini 2.5 Pro & \hspace{.5em}9.2 & 0.81 & \textcolor{green!60!black}{14.4}\hspace{.5em}\textcolor{lightgray}{(1.6{\scriptsize$\times$)}} & 1.16  \\
Qwen 3 Max & 24.4 & 4.32 & \textcolor{green!60!black}{35.8}\hspace{.5em}\textcolor{lightgray}{(1.5{\scriptsize$\times$)}} & 5.67  \\
DeepSeek V3.2 & 38.9 & 0.08 & \textcolor{red!80!black}{10.5}\hspace{.5em}\textcolor{lightgray}{(0.3{\scriptsize$\times$)}} & 0.15  \\
\cmidrule(r){1-1}
\cmidrule(lr){2-2}
\cmidrule(lr){3-3}
\cmidrule(lr){4-4}
\cmidrule(l){5-5}
\textbf{Mean} & 17.2 & 0.80 & \textcolor{green!60!black}{20.3}\hspace{.5em}\textcolor{lightgray}{(1.2{\scriptsize$\times$})} & 1.05  \\
    \bottomrule
    \end{tabular}
\end{table}

\begin{figure}[tb]
    \centering
    \includegraphics[width=\linewidth]{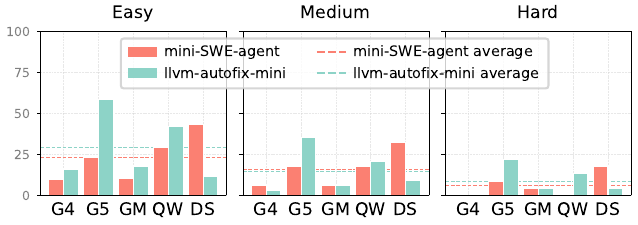}
    \caption{
        \textbf{\ouragent outperforms \mswe at each split, but as splits increase in difficulty, frontier models tend to struggle or fail}.
        The Y-axis is ``\%Passed''.
        \code{G4}, \code{G5}, \code{GM}, \code{QW}, and \code{DS} are short for GPT 4o, GPT 5, Gemini 2.5 Pro, Qwen 3 Max, and DeepSeek V3.2.
    }
    \label{fig:mswe-ours-results}
\end{figure}

\subsection{Baseline Comparison and Common Failures}\label{sec:rq3}
\begin{figure}[tb]
    \centering
    \includegraphics[width=\linewidth]{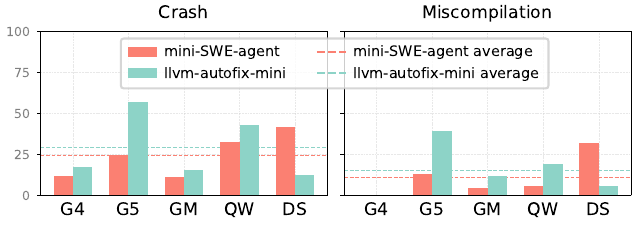}
    \caption{
        \textbf{\ouragent outperforms \mswe for each bug type, and miscompilations are more challenging than crashes}.
        The Y-axis is ``\%Passed''.
        \code{G4}, \code{G5}, \code{GM}, \code{QW}, and \code{DS} are short for GPT 4o, GPT 5, Gemini 2.5 Pro, Qwen 3 Max, and DeepSeek V3.2.
    }
    \label{fig:results-by-bug-type}
\end{figure}

Finally, we compare the performance of \ouragent with \ourplatform-enhanced \mswe.
As shown in \Cref{tab:mswe-ours-results}, \ouragent's performance is generally 1.22$\times$ of \mswe's with affordable cost.
A statistical test confirming \ouragent's significant improvement is presented in \Cref{sec:deeper-analysis} of our appendix.
The greatest improvement occurs with GPT 5 (+145\%), whereas DeepSeek V3.2 is an exception (-73\%):
when executed with \ouragent, it frequently fails to adhere to our tool-calling format, whereas such failures are seldom seen with \mswe;
this results in a large (>85\%) portion of failures exceeding the token budget.%
\footnote{
Note that this is a commonly encountered, DeepSeek V3.2-specific issue reported in other settings as well~\cite{dsv32issue1,dsv32issue2,dsv32issue3,dsv32issue4}, rather than an issue of our format.
}
The improvement for GPT 4o, Gemini 2.5 Pro, and Qwen 3 Max is 47\%, 57\%, and 47\%, respectively.
For all agent-model pairs, GPT 5 successfully resolves the largest number of issues (118/229, 51.5\%) when executed with \ouragent, followed by DeepSeek V3.2 with \mswe (89/229, 38.9\%) and Qwen 3 Max with \ouragent (82/229, 35.8\%).
The bug repair capability of the four frontier models is generally better than the baseline model: improved by around 16\% when executed with \ouragent, compared with GPT 4o.
The improvement of the GPT, Qwen, and DeepSeek models is more marked than that of the Gemini model.
The same trend, i.e., \ouragent outperforming \mswe, is consistently observed across different difficulty splits (\Cref{fig:mswe-ours-results}) and bug types (\Cref{fig:results-by-bug-type}).

\smalltitle{By difficulty}
The resolution rate decreases as splits become more difficult for both agents.
The \code{hard} split poses a significant challenge (\Cref{fig:mswe-ours-results}):
Every selected model can fix certain bugs in the \code{easy} or \code{medium} split, but in the \code{hard} split, \ouragent cannot drive GPT 4o to repair any issue.
In particular, \mswe's average resolution rates are 23.2\%, 15.8\%, and 6.1\% for \code{easy}, \code{medium}, and \code{hard}, respectively, whereas the data for \ouragent are 28.8\% (1.8$\times$), 14.7\% (0.9$\times$), and 7.8\% (1.3$\times$).
The degradation for the \code{medium} split is primarily due to DeepSeek V3.2 as described earlier.
We believe that this indicates the necessity of developing LLVM-specific auto-fixing agents with \ourplatform, rather than relying on a general-purpose agent scaffold such as \mswe which targets common software.
\ouragent serves as a more promising baseline.

\smalltitle{By bug type}
\Cref{fig:results-by-bug-type} illustrates both agents' performance breakdown across different bug types.
Generally, miscompilations are more challenging to diagnose and repair than crashes, which aligns with our experience.
The resolution rate of miscompilations lags approximately 13\% and 14\% behind crashes for \ouragent and \mswe, respectively.
The baseline model GPT 4o is unable to resolve any miscompilations, whether using \mswe or \ouragent.
Among frontier models, GPT 5 and DeepSeek V3.2 are the best-performing models for \ouragent and \mswe, respectively, whereas Gemini 2.5 Pro's capability usually falls below average for either bug type.

\begin{figure}[tb]
    \includegraphics[width=\linewidth]{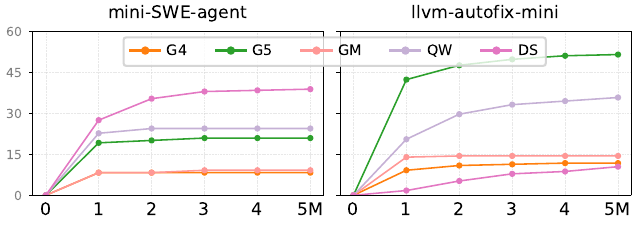}
    \caption{
        \textbf{\ouragent is more robust as token limit increases than \mswe}.
        The Y-axis is ``\%Passed''.
        For \mswe, nearly all models converge after reaching three million tokens.
        \code{G4}, \code{G5}, \code{GM}, \code{QW}, and \code{DS} are short for GPT 4o, GPT 5, Gemini 2.5 Pro, Qwen 3 Max, and DeepSeek V3.2.
    }
    \label{fig:results-with-token-limit}
\end{figure}

\smalltitle{With token limit}
We investigate whether the agent's performance improves as token limits increase.
\Cref{fig:results-with-token-limit} displays this trend.
\mswe's performance appears to converge when the token limit reaches three million.
Although \ouragent shows a similar pattern with GPT 4o and Gemini 2.5 Pro, \ouragent is more robust:
the performance of the other three frontier models increases smoothly.
Aside from DeepSeek V3.2, \ouragent consistently outperforms \mswe once the token budget exceeds two million and continues to improve until the five-million budget.
However, for the one-million token budget, \ouragent performs slightly worse than Qwen 3 Max due to the additional reasoning stage required before attempting fixes.
It is also interesting to note that the performance of DeepSeek V3.2 increases more markedly than that of other models when executed with \ouragent, and it is expected to surpass GPT 4o as the limit extends to six million tokens.
An explanation for this could be that DeepSeek V3.2 progressively adheres to our tool calling format as more interactions are involved.

On the other hand, the performance of these models does \emph{not} notably improve with an increased number of tokens.
We propose two hypotheses in this paper:
(1) The ``context rot'' problem~\cite{contextrot} becomes more pronounced as agents operate for extended periods (resulting in increased context).
(2) The complexity of compiler bugs makes it increasingly infeasible to resolve the remaining issues within a reasonable token limit.

\begin{table}[tb]
    \centering
    \scriptsize
    \setlength{\tabcolsep}{.5em}
    \caption{
        Cost and Overhead of \ouragent and \mswe.
        M: Model;
        T/m: Time/min;
        \#Tot/\#In/\#Out/\#Cache: The total/input/output/cached tokens consumed (measured in millions unless otherwise specified);
        \code{G4}, \code{G5}, \code{GM}, \code{QW}, and \code{DS}: GPT 4o, GPT 5, Gemini 2.5 Pro, Qwen 3 Max, and DeepSeek V3.2.
    }
    \label{tab:cost-overhead}
    \begin{tabular}{lccccccccc}
        \toprule
        & & & & \multicolumn{5}{c}{\bf Token Cost (million)} & \\ 
        \cmidrule(lr){5-9}
        \bf M & \bf T/m & \bf \#Tests & \bf \#Rounds & \it \#Tot & \it \#In & \it \#Out & \it \#Cache & \it \%Cache & \bf \$Cost \\
\midrule
\multicolumn{10}{c}{\cellcolor{lightgray!40!white}\code{mini-SWE-agent}\vspace*{.5em}}\\
\code{G4} & \hspace{.5em}7.1 & 13.0 & \hspace{.5em}74.4 & 2.2 & 2.2 & \hspace{.5em}6.9K & 2.2 & 96.4 & \hspace{.5em}1.6 \\
\code{G5} & 14.6 & \hspace{.5em}9.5 & \hspace{.5em}66.9 & 2.6 & 2.6 & 28.2K & 2.5 & 94.4 & \hspace{.5em}0.7 \\
\code{GM} & \hspace{.5em}8.8 & \hspace{.5em}7.4 & \hspace{.5em}51.7 & 1.6 & 1.6 & 21.3K & 1.4 & 85.8 & \hspace{.5em}0.8 \\
\code{QW} & 18.1 & 13.1 & \hspace{.5em}77.7 & 2.4 & 2.4 & 15.3K & 1.1 & 43.9 & \hspace{.5em}4.3 \\
\code{DS} & 15.1 & \hspace{.5em}3.9 & \hspace{.5em}80.0 & 2.3 & 2.3 & 17.2K & 2.2 & 97.9 & \hspace{.5em}0.1 \\
\midrule
\multicolumn{10}{c}{\cellcolor{lightgray!40!white}\code{llvm-autofix-mini}\vspace*{.5em}}\\
\code{G4} & 15.5 & 10.4 & 116.5 & 3.1 & 3.1 & \hspace{.5em}7.7K & 3.0 & 96.3 & \hspace{.5em}2.2 \\
\code{G5} & 16.1 & \hspace{.5em}6.4 & \hspace{.5em}39.0 & 1.3 & 1.3 & 32.3K & 1.2 & 90.0 & \hspace{.5em}0.6 \\
\code{GM} & 12.5 & \hspace{.5em}3.5 & \hspace{.5em}53.2 & 1.5 & 1.5 & 21.5K & 1.3 & 83.3 & \hspace{.5em}1.2 \\
\code{QW} & 25.5 & \hspace{.5em}8.0 & \hspace{.5em}81.4 & 3.4 & 3.4 & 16.8K & 1.7 & 49.3 & \hspace{.5em}5.7 \\
\code{DS} & 21.8 & \hspace{.5em}1.3 & 183.5 & 4.7 & 4.7 & 16.3K & 4.6 & 98.5 & \hspace{.5em}0.2 \\
        \bottomrule
    \end{tabular}
\end{table}

\smalltitle{Cost and overhead}
Fixing LLVM issues necessitates more cost and runtime overhead compared with common software issues in \swev (\Cref{tab:mswe-results}).
The average overhead (\Cref{tab:cost-overhead}) for resolving an issue using the four frontier models with \mswe is 14.1 minutes, ranging from 7.1 to 18.1 minutes.
While \ouragent requires an additional 4.8 minutes on average, this overhead remains practical.
Additionally, the costs for both agents are highly affordable (maximum \$6/issue);
for their best-performing models, the cost is as low as \$0.1 and \$0.6 per issue, respectively.
We also observe that \ouragent typically engages in more rounds of reasoning and information gathering, while requiring fewer \lstcode{test} calls compared to \mswe.
This suggests that our design enables models to make more informed and efficient decisions.
Our agent is as cache-friendly as \mswe, enabling it to fully leverage the token budget.

\begin{figure}[tb]
    \centering
    \includegraphics[width=\linewidth]{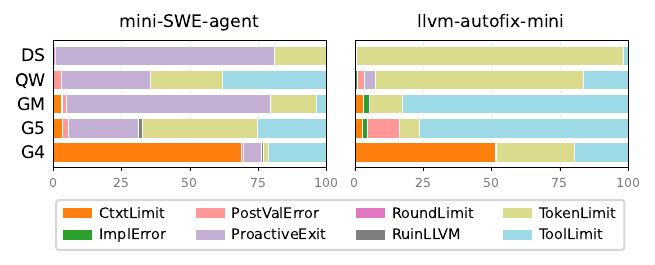}
    \caption{
        Failure distribution of unresolved issues.
        \code{G4}, \code{G5}, \code{GM}, \code{QW}, and \code{DS} are short for GPT 4o, GPT 5, Gemini 2.5 Pro, Qwen 3 Max, and DeepSeek V3.2, respectively.
    }
    \label{fig:failure-distribution}
\end{figure}

\smalltitle{Common LLM failures}
We group failures by LLMs in \Cref{fig:failure-distribution}.
GPT 4o typically fails due to context limitation (\code{CtxtLimit}): it frequently calls tools such as \lstcode{read} whose output is long, exceeding the 64K-token context window.
The four frontier models rarely have such failures, possibly due to their stronger understanding capability.
Instead, their failure symptoms vary depending on the underlying agent.
For \ouragent, the majority of failures are \code{TokenLimit} and \code{ToolLimit}, i.e., reaching the limit of the token budget (5M tokens) or tool budget (25 \lstcode{edit}s or 25 \lstcode{test}s).
If given a larger budget, we expect \ouragent to continue executing.
Although \code{TokenLimit} and \code{ToolLimit} account for nearly half as well, a large number of \mswe failures are \code{ProactiveExit}
where the agent proactively exits the repair process, declaring that it cannot fix the issue.
There are cases where \mswe corrupts the LLVM repository (\code{RuinLLVM}), leading to crashes during patch validation;
such failures do not occur with \ouragent.
Post validation failures (\code{PostValError}) take up to 12\%.
Two bugs in \ouragent caused several \code{ImplError}s.
We did not observe \code{RoundLimit} errors.

\begin{table}[tb]
    \centering
    \scriptsize
    \setlength{\tabcolsep}{.9em}
    \caption{
        \textbf{Current LLMs lack satisfactory, true capability in fixing LLVM middle-end issues}:
        Despite passing all tests, fewer than 42\% of passed patches are valid.
        ``\%Valid = \#ValidPatches / \#Issues'' is the percentage of valid patches.
    }
    \label{tab:human-study-results}
    \begin{tabular}{lcccc}
    \toprule
    & \multicolumn{2}{c}{\mswe} & \multicolumn{2}{c}{\ouragent} \\
    \cmidrule(lr){2-3} \cmidrule(l){4-5}
    \bf Model & \bf \%Passed & \bf \%Valid & \bf \%Passed & \bf \%Valid \\
    \cmidrule(r){1-1}
    \cmidrule(lr){2-2}
    \cmidrule(lr){3-3}
    \cmidrule(lr){4-4}
    \cmidrule(l){5-5}
GPT 4o & \hspace{.5em}8.3 & \textcolor{red}{\hspace{.5em}1.7} \textcolor{lightgray}{(21.0{\scriptsize\%})} & 12.2 & \textcolor{red}{\hspace{.5em}3.9} \textcolor{lightgray}{(32.1{\scriptsize\%})} \\
\cmidrule(r){1-1}
\cmidrule(lr){2-2}
\cmidrule(lr){3-3}
\cmidrule(lr){4-4}
\cmidrule(l){5-5}
GPT 5 & 21.0 & \textcolor{red}{\hspace{.5em}6.6} \textcolor{lightgray}{(31.3{\scriptsize\%})} & 51.5 & \textcolor{red}{20.1} \textcolor{lightgray}{(39.0{\scriptsize\%})} \\
Gemini 2.5 Pro & \hspace{.5em}9.2 & \textcolor{red}{\hspace{.5em}2.2} \textcolor{lightgray}{(23.8{\scriptsize\%})} & 14.4 & \textcolor{red}{\hspace{.5em}5.2} \textcolor{lightgray}{(36.4{\scriptsize\%})} \\
Qwen 3 Max & 24.4 & \textcolor{red}{\hspace{.5em}8.7} \textcolor{lightgray}{(35.7{\scriptsize\%})} & 35.8 & \textcolor{red}{13.1} \textcolor{lightgray}{(36.6{\scriptsize\%})} \\
DeepSeek V3.2 & 38.9 & \textcolor{red}{14.4} \textcolor{lightgray}{(37.1{\scriptsize\%})} & 10.5 & \textcolor{red}{\hspace{.5em}4.4} \textcolor{lightgray}{(41.7{\scriptsize\%})} \\
\cmidrule(r){1-1}
\cmidrule(lr){2-2}
\cmidrule(lr){3-3}
\cmidrule(lr){4-4}
\cmidrule(l){5-5}
\textbf{Mean} & 20.3 & \textcolor{red}{\hspace{.5em}6.7} \textcolor{lightgray}{(29.8{\scriptsize\%})} & 24.9 & \textcolor{red}{\hspace{.5em}9.3} \textcolor{lightgray}{(37.1{\scriptsize\%})} \\
    \bottomrule
    \end{tabular}
\end{table}

\begin{figure}[tb]
    \centering
    \includegraphics[width=\linewidth]{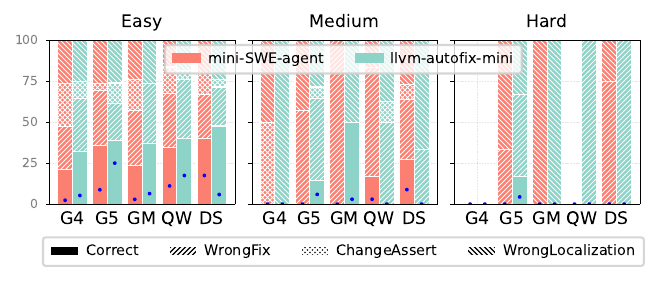}
    \caption{
        \textbf{As splits increase in difficulty, frontier models' true capability degrades fast}:
        No models can handle the \code{hard} split except GPT 5 (\ouragent).
        \code{G4}, \code{G5}, \code{GM}, \code{QW}, and \code{DS} are short for GPT 4o, GPT 5, Gemini 2.5 Pro, Qwen 3 Max, and DeepSeek V3.2, respectively.
        Blue bullets (\textcolor{blue}{\Large{$\cdot$}}) represent the rate of truly resolved issues, i.e., ``\%Valid'' in \Cref{tab:human-study-results}.
    }
    \label{fig:human-study-results-by-difficulty}
\end{figure}


\section{Expert Review Study}\label{sec:rq4}

Although passing all tests, passed patches may still be invalid due to the limitation of dynamic testing.
Therefore, to better understand the \emph{true} capability of frontier models,
we employ an expert, who is an active LLVM developer and maintainer, to review every passed patch from \mswe and \ouragent.
Although involving more experts would reduce bias and strengthen the study,
review capacity is a scarce resource in the LLVM community.
The project itself therefore adopts a single-reviewer model,
where a patch can be committed once one reviewer approves it~\cite{llvmreviewprinciple}.
We follow this established practice.
The expert labels \emph{valid patches} (semantically sound and review-worthy for LLVM developers) and categorizes common LLM mistakes among invalid ones.
That said, in this study, we recognize an issue as being \emph{truly resolved} if the passed patch is labeled valid by the expert reviewer.
\Cref{tab:human-study-results} shows the percentage of valid patches after code review (``\%Valid = \#ValidPatches / \#Issues''), as well as their proportion among passed patches (in gray).
Our interpretation of the review result is twofold.

First, we observe that \ouragent's advantage becomes even more pronounced after code review:
(1) it produces 38\% more valid patches than \mswe, up from a 22\% advantage before code review, and
(2) the percentage of valid patches among benchmark-passing patches is 24\% higher compared to \mswe.
Our best-performing model, GPT 5, surpasses \mswe's best-performing model DeepSeek V3.2 by 40\%.
Similar trends are also observed in each split (\Cref{fig:human-study-results-by-difficulty}) and bug type (\Cref{tab:human-study-by-bug-type}).
The expert even found instances where the \ouragent-provided patch is better than the golden patch provided by LLVM developers;
such an example is present in the appendix (\Cref{sec:example-patches}).

Second, however, we find that \emph{truly resolving LLVM middle-end issues is exceptionally difficult}.
Despite advances over the baseline model GPT 4o, frontier models still do not possess adequate, true capabilities for repairing LLVM middle-end issues.
Even for the best-performing agent \ouragent (GPT 5), more than 60\% of passed patches are invalid, resulting in only 20.1\% of issues being truly resolved;
all other agent-model pairs are consistently below 15\%, where 7 out of 10 agent-models are below 9\% (21/229).
The results become even worse as the split difficulty increases (\Cref{fig:human-study-results-by-difficulty}).
Many agent-model pairs (6/10) are unable to output any valid patches for the \code{medium} split.
In terms of the \code{hard} split, only GPT 5 manages to resolve a single issue with \ouragent.
This \%Passed-\%Valid gap arises in all LLMs because LLVM tests typically capture specific symptoms, rather than a complete compilation specification.
When comparing crashes and miscompilations (\Cref{tab:human-study-by-bug-type}), the rate of valid miscompilation patches is generally lower than that of valid crash patches for both agents, aligning with the observation that miscompilations are more challenging.

\begin{table}[tb]
    \centering
    \scriptsize
    \setlength{\tabcolsep}{.5em}
    \caption{
        \textbf{Miscompilations are more challenging than crashes}.
        Valid miscompilation patches are usually fewer than valid crash patches.
        ``\code{miscomp}'' stands for miscompilation.
    }
    \begin{tabular}{llcccc}
    \toprule
    & & \multicolumn{2}{c}{\mswe} & \multicolumn{2}{c}{\ouragent} \\
    \cmidrule(lr){3-4} \cmidrule(l){5-6}
    \bf Type & \bf Model & \bf \%Passed & \bf \%Valid & \bf \%Passed & \bf \%Valid \\
    \cmidrule(r){1-1}
    \cmidrule(r){2-2}
    \cmidrule(lr){3-3}
    \cmidrule(lr){4-4}
    \cmidrule(l){5-5}
    \cmidrule(l){6-6}
\multirow{6}{*}{\code{crash}} & GPT 4o & 11.9 & \textcolor{red}{\hspace{.5em}2.5} \textcolor{lightgray}{(21.1{\scriptsize\%})} & 17.5 & \textcolor{red}{\hspace{.5em}5.6} \textcolor{lightgray}{(32.1{\scriptsize\%})} \\
\cmidrule(r){2-2}
\cmidrule(lr){3-3}
\cmidrule(lr){4-4}
\cmidrule(l){5-5}
\cmidrule(l){6-6}
& GPT 5 & 24.4 & \textcolor{red}{\hspace{.5em}6.9} \textcolor{lightgray}{(28.2{\scriptsize\%})} & 56.9 & \textcolor{red}{18.1} \textcolor{lightgray}{(31.9{\scriptsize\%})} \\
& Gemini 2.5 Pro & 11.2 & \textcolor{red}{\hspace{.5em}2.5} \textcolor{lightgray}{(22.2{\scriptsize\%})} & 15.6 & \textcolor{red}{\hspace{.5em}6.2} \textcolor{lightgray}{(40.0{\scriptsize\%})} \\
& Qwen 3 Max & 32.5 & \textcolor{red}{11.2} \textcolor{lightgray}{(34.6{\scriptsize\%})} & 43.1 & \textcolor{red}{15.0} \textcolor{lightgray}{(34.8{\scriptsize\%})} \\
& DeepSeek V3.2 & 41.9 & \textcolor{red}{11.9} \textcolor{lightgray}{(28.4{\scriptsize\%})} & 12.5 & \textcolor{red}{\hspace{.5em}4.4} \textcolor{lightgray}{(35.0{\scriptsize\%})} \\
\cmidrule(r){2-2}
\cmidrule(lr){3-3}
\cmidrule(lr){4-4}
\cmidrule(l){5-5}
\cmidrule(l){6-6}
& \textbf{Mean} & 21.3 & \textcolor{red}{\hspace{.5em}5.6} \textcolor{lightgray}{(26.5{\scriptsize\%})} & 24.2 & \textcolor{red}{\hspace{.5em}8.4} \textcolor{lightgray}{(34.6{\scriptsize\%})}  \\
\cmidrule(r){1-1}
\cmidrule(r){2-2}
\cmidrule(lr){3-3}
\cmidrule(lr){4-4}
\cmidrule(l){5-5}
\cmidrule(l){6-6}
\multirow{6}{*}{\code{miscomp}} & GPT 4o & \hspace{.5em}0.0 & \textcolor{red}{\hspace{.5em}0.0} \textcolor{lightgray}{(\hspace{.5em}0.0{\scriptsize\%})} & \hspace{.5em}0.0 & \textcolor{red}{\hspace{.5em}0.0} \textcolor{lightgray}{(\hspace{.5em}0.0{\scriptsize\%})} \\
\cmidrule(r){2-2}
\cmidrule(lr){3-3}
\cmidrule(lr){4-4}
\cmidrule(l){5-5}
\cmidrule(l){6-6}
& GPT 5 & 13.0 & \textcolor{red}{\hspace{.5em}5.8} \textcolor{lightgray}{(44.5{\scriptsize\%})} & 39.1 & \textcolor{red}{24.6} \textcolor{lightgray}{(63.0{\scriptsize\%})} \\
& Gemini 2.5 Pro & \hspace{.5em}4.3 & \textcolor{red}{\hspace{.5em}1.4} \textcolor{lightgray}{(33.3{\scriptsize\%})} & 11.6 & \textcolor{red}{\hspace{.5em}2.9} \textcolor{lightgray}{(25.0{\scriptsize\%})} \\
& Qwen 3 Max & \hspace{.5em}5.8 & \textcolor{red}{\hspace{.5em}2.9} \textcolor{lightgray}{(50.0{\scriptsize\%})} & 18.8 & \textcolor{red}{\hspace{.5em}8.7} \textcolor{lightgray}{(46.2{\scriptsize\%})} \\
& DeepSeek V3.2 & 31.9 & \textcolor{red}{20.3} \textcolor{lightgray}{(63.6{\scriptsize\%})} & \hspace{.5em}5.8 & \textcolor{red}{\hspace{.5em}4.3} \textcolor{lightgray}{(75.0{\scriptsize\%})} \\
\cmidrule(r){2-2}
\cmidrule(lr){3-3}
\cmidrule(lr){4-4}
\cmidrule(l){5-5}
\cmidrule(l){6-6}
& \textbf{Mean} & 10.1 & \textcolor{red}{\hspace{.5em}4.7} \textcolor{lightgray}{(46.6{\scriptsize\%})} & 14.9 & \textcolor{red}{\hspace{.5em}7.2} \textcolor{lightgray}{(48.3{\scriptsize\%})}  \\
    \bottomrule
    \end{tabular}
    \label{tab:human-study-by-bug-type}
\end{table}


\smalltitle{Common LLM mistakes}
The expert identifies three categories of LLM mistakes (examples in \Cref{sec:example-patches}):

\begin{itemize}
    \item \code{ChangeAssert}.
    LLMs ``cheat'' the system by altering the assertion condition to avoid assertion failures (one type of crash bugs).
    Their erroneous code changes include
    (1) changing the assertion condition itself, 
    (2) modifying functions directly or indirectly invoked by the assertion condition, and
    (3) inserting early return statements preceding the assertion.

    \item \code{WrongLocalization}.
    Despite \ourplatform providing \ouragent and \mswe with the erroneous component, models still struggle to localize bugs within erroneous files and functions, particularly when bugs impact multiple of them.
    Detailed data regarding bug localization is presented in \Cref{sec:localization} of the appendix in our supplementary material.

    \item \code{WrongFix}.
    Models generate passed yet faulty patches although with correct localization.
    However, for LLVM and other compilers, such patches often suffer from several limitations:
    (1) they bypass the erroneous component by weakening or strengthening the activation condition instead of addressing the issue,
    (2) they lack generality beyond the specific reproducers and regression tests, and
    (3) they silently introduce new bugs, such as missed optimizations that inhibit subsequent optimizations.
    Although this does not affect correctness, it negatively affects the performance of generated code.
\end{itemize}

\section{Discussion}\label{sec:discussion}

In summary, our benchmarking experiments and expert review study of frontier models highlight a key finding:
\begin{center}
\emph{Truly resolving LLVM middle-end issues in \ourbenchmark is exceptionally challenging, where specialized harnesses and tailored agents are required}.
\end{center}
In this context, \ourplatform and \ouragent lay the foundations for the future;
see more in \Cref{sec:llm-for-compilers}.



\smalltitle{Threats to validity}
The first threat concerns our use of a single sample to report pass@1.
Due to budget limitations, we cannot run multiple trials.
We mitigate this threat by using greedy decoding for each model call, a standard choice in the community.
Second, there is a potential threat of data contamination.
To examine its influence, we compare the resolution rate of issues that were fixed before and after the release of each model.
Our results show that data contamination does not appear to have a substantial impact, especially on \ouragent.
We include the detailed results of this study in \Cref{sec:data-leakage}.
Finally, our expert review findings may render ``\%Passed'' used in the benchmarking experiment less useful.
However, we argue that it remains valuable because
it is objective, reproducible, and fully automated.
It indicates an upper bound of each LLM's capability and
allows sound comparison across LLMs.
For example, \Cref{tab:human-study-results} shows that
the relative ranking of LLMs' performance remains unchanged
after expert review, for both agents.

\smalltitle{Limitations and extensions}
Our expert review study in \Cref{sec:rq4} implies that \ourplatform is currently limited by LLVM's regression tests, leading to a \%Passed-\%Valid gap.
These tests are inadequate to systematically validate agent-generated patches, especially in detecting the three categories of LLM mistakes.
However, unlike common software, addressing such issues poses unique challenges for compilers;
we discuss this later in Section~\ref{sec:llm-for-compilers}.

\ourplatform does not currently consider performance issues related to the middle-end, such as slow compilations, compiler hangs, and missed optimizations, nor does it include frontend and backend bugs.
In fact, we have already incorporated compiler hangs by the time of submitting this paper.
Extending \ourplatform to the backend requires additional engineering effort, such as integrating
backend tooling including \code{llc} for backend compilation and optimization,
target-dependent \code{arm-tv} and \code{riscv-tv} for translation validation, and
mid–backend cross-validation with \code{llubi} and \code{llc}.
The same applies to various LLVM frontends.
Our agent \code{llvm-autofix-mini}, however, is end-independent and could be adapted by wiring it to the corresponding toolsets.
We are now working on these extensions.

\section{Open Challenges for Future Work}\label{sec:llm-for-compilers}

Although outperforming \mswe, the expert review study exposes several challenges that warrant further advances to create practical, more effective LLVM auto-fixing agents, along two directions.

\smalltitle{Patch generation}
\ouragent achieves at best a resolution rate of only 51.5\%,
with an average of 24.9\% across settings (\Cref{sec:rq1}),
highlighting significant room for improvement.
Future work should prioritize the development of more advanced agents on patch generation to consistently achieve higher \%Passed scores,
particularly for medium and hard splits as well as miscompilation bugs.  
For instance, many of our failures stem from token/tool limitations (\Cref{sec:rq3}) and incorrect bug localization (\Cref{sec:rq4}).
Enhancing agents' expertise in LLVM-related tasks should therefore be a primary focus;
a promising avenue is to distill knowledge and bug-fixing experiences from LLVM resources and developers into reusable skills.
Instead of relying on static workflows (\Cref{sec:agent}), next-generation agents may benefit from adopting more flexible, autonomous designs that dynamically adapt to varying levels of task difficulty.
Compilers' complexity necessitates large context limits (\Cref{sec:rq3}), but long context can lead to the ``context rot'' problem~\cite{contextrot}.
Designing sophisticated context management mechanisms that maintain concise yet tractable context is also important.

\smalltitle{Patch validation}
\Cref{sec:rq4} shows there exists a \%Passed-\%Valid gap, underscoring the limitations of relying solely on LLVM tests.
It is crucial to develop automated patch validation techniques or agents because relying on experts' manual review to reduce the gap is impractical.
First, LLMs can manipulate logic by altering conditions or inserting shortcuts that bypass essential checks or optimizations.
While such bypasses might lead to errors in general software, in the context of compilers, they result in technically \emph{correct yet suboptimal} code, a \emph{unique challenge} in the compiler domain.
Patch validation techniques should discover such bypasses, and further detect such missed optimizations.
Second, LLMs tend to stop repairs when current tests are passed.
Such ``oversight'' often results in overfitting patches that fail to optimize unseen code, leading to new bugs.
Patch validation techniques should be able to handle these patches.

%% file: sections/relatedwork.tex
\section{Related Work}


\smalltitle{Compiler testing}
Recent research has increasingly leveraged LLMs to enhance compiler validation and verification.
Some works integrate LLMs into the fuzzing process to generate random input programs~\cite{fuzz4all,whitefox,legofuzz,llm4vv} and mutation operators~\cite{metamut}, targeting crash bugs and miscompilation bugs.
Other works focus on detecting performance issues such as missed-optimizations~\cite{missedsizeopt,missedpeepholeopt}.
These works demonstrate significant potential, and \ourplatform offers an opportunity to address the bugs that they uncover automatically through frontier models and our agents.

\smalltitle{Compiler optimization}
Integrating traditional ML techniques and LLMs into the compilation pipeline also shows promise for improving both code size and performance.
These works usually target specific compiler components, such as
inlining~\cite{mlgo,llmcompiler},
register allocation~\cite{rl4real,llmcompiler},
loop optimization~\cite{llmvec},
pass scheduling~\cite{phraseorderjikes},
auto-tuning~\cite{autotunesurvey}, and
super optimization~\cite{supercode,llmrlcodeperf}.
In this context, CompilerGym provides a toolkit for applying reinforcement learning to compiler optimization tasks~\cite{compilergym}.
Unlike ours, their focus is on improving the optimizations themselves rather than addressing reported bugs within them.

\smalltitle{Other systems tasks}
Beyond compilers, LLMs have been applied to other systems tasks.
For instance, certain studies focus on Linux kernel crash diagnosis and repair, offering benchmarks and environments to evaluate LLM-based agents~\cite{kgym,crashfixer,rgym}.
Additionally, research has explored LLMs' ability to synthesize efficient GPU kernels~\cite{kernelbench}.

%% file: sections/conclusion.tex
\section{Conclusion}


Despite LLMs' success in common software, they face challenges with compiler-specific issues.
\ourplatform offers essential LLVM-specific tools and a challenging benchmark, promoting the development of specialized agents like \ouragent.
Our expert reviews also highlight ongoing challenges, including improving model expertise and devising robust patch validation methods.

%% file: sections/implementation.tex
\section{More Implementation Details}

\subsection{Affected Middle-End Components}\label{sec:full-list-affected-componnets}

{SLPVectorizer} (79),
{LoopVectorize} (79),
{InstCombine} (54),
{ScalarEvolution} (16),
{VectorCombine} (13),
{ValueTracking} (9),
{IR} (7),
{ConstraintElimination} (6),
{InstructionSimplify} (5),
{SimplifyIndVar} (4),
{MemorySSAUpdater} (4),
{LoopPeel} (4),
{LoopAccessAnalysis} (3),
{Local} (3),
{GVN} (3),
{FunctionAttrs} (3),
{ConstantFold} (3),
{SimplifyCFG} (2),
{LoopStrengthReduce} (2),
{LoopSimplifyCFG} (2),
{LazyValueInfo} (2),
{LICM} (2),
{GlobalOpt} (2),
{EarlyCSE} (2),
{DeadStoreElimination} (2),
{VectorUtils} (1),
{ValueMapper} (1),
{SimplifyLibCalls} (1),
{SimpleLoopUnswitch} (1),
{SeparateConstOffsetFromGEP} (1),
{Scalarizer} (1),
{SCCPSolver} (1),
{SCCP} (1),
{Reassociate} (1),
{PredicateInfo} (1),
{NewGVN} (1),
{MoveAutoInit} (1),
{MemCpyOptimizer} (1),
{LowerSwitch} (1),
{LoopUnrollRuntime} (1),
{LoopUnrollAndJamPass} (1),
{LoopFuse} (1),
{LoopDeletion} (1),
{LoopCacheAnalysis} (1),
{Loads} (1),
{JumpThreading} (1),
{Instrumentation} (1),
{InlineCost} (1),
{InductiveRangeCheckElimination} (1),
{IndVarSimplify} (1),
{GVNSink} (1),
{FunctionSpecialization} (1),
{Evaluator} (1),
{DemoteRegToStack} (1),
{DeadArgumentElimination} (1),
{DFAJumpThreading} (1),
{CorrelatedValuePropagation} (1),
{Coroutines} (1),
{BreakCriticalEdges} (1),
{BasicBlockUtils} (1),
{BDCE} (1),
{Attributor} (1),
{AliasAnalysis} (1),
{AggressiveInstCombine} (1).

\subsection{Enabled Tools in Benchmarking Experiment}\label{sec:enabled-tools-in-eval}

\begin{table}[h]
    \scriptsize
    \centering
    \renewcommand{\code}[1]{\texttt{{#1}}}
    \renewcommand{\lstcode}[1]{\lstinline[basicstyle=\ttfamily]|#1|}

    
    \begin{tabular}{ll}
      \toprule
      \bf Signature & \bf Description \\
      \midrule
\rowcolor{gray!15!white}
\multicolumn{2}{c}{Generic Tools}\vspace{.2em} \\
\lstcode{find(pat,dir)} & Find files matching the pattern \code{pat} in the directory \code{dir} recursively \\
\lstcode{grep(args)} & Find text matching patterns and files mentioned in arguments \code{args} \\
\lstcode{list(dir)} & List files and directories in the directory \code{dir} \\
\lstcode{read(file,pos,k)} & Read \code{k} lines of content starting from line \code{pos} of the file \code{file} \\
\lstcode{edit(file,text,rep)} & Replace the text \code{text} with the text \code{rep} in the file \code{file} \\
\lstcode{preview()} & Preview LLM's code changes in the format of unified diff \\
\midrule
\rowcolor{gray!15!white}
\multicolumn{2}{c}{LLVM-specific Tools}\vspace{.2em} \\
\lstcode{code(func)} & List the code of the LLVM function \code{func} \\
\lstcode{docs(func)} & Show the documentation of the LLVM function \code{func} \\
\lstcode{langref(inst)} & Show the language reference of the LLVM IR instruction \code{inst} \\
\lstcode{debug(cmd)} & Run GDB command \code{cmd} over the current LLVM interal state \\
\lstcode{eval(expr)} & Evaluate the LLVM internal state by the expression \code{expr} \\
\lstcode{reset()} & Reset the LLVM project to its initial state of this bug \\
\lstcode{test()} & Test the LLM's patch with LLVM online and return the feedback \\
      \bottomrule
    \end{tabular}
\end{table}

\subsection{Other Available Tools}\label{sec:other-harness-tools}

\begin{table}[h]
    \scriptsize
    \centering
    \renewcommand{\code}[1]{\texttt{{#1}}}
    \renewcommand{\lstcode}[1]{\lstinline[basicstyle=\ttfamily]|#1|}

    
    \begin{tabular}{ll}
      \toprule
      \bf Signature & \bf Description \\
      \midrule
\rowcolor{gray!15!white}
\multicolumn{2}{c}{Generic Tools}\vspace{.2em} \\
\lstcode{write(file,txt)} & Write \code{txt} into \code{file} \\
\lstcode{bash(cmd,tmo)} & Execute command \code{cmd} in a Bash environment with \code{tmo} as timeout \\
\lstcode{todo(act,args)} & Perform a TODO action \code{act} with \code{args} \\
\midrule
\rowcolor{gray!15!white}
\multicolumn{2}{c}{LLVM-specific Tools}\vspace{.2em} \\
\lstcode{build()} & Build a debuggable LLVM \\
\lstcode{interp(file,args)} & Interpret LLVM IR program \code{file} with \code{args} via \code{llubi} (or \code{lli} as a fallback) \\
\lstcode{opt(src,tgt,args)} & Optimize LLVM IR program \code{src} into \code{tgt} via \code{opt} \\
\lstcode{exec(file,args)} & Execute LLVM IR program \code{file} with \code{args} via \code{lli} \\
\lstcode{comp(src,tgt,args)} & Compile LLVM IR program \code{src} into machine code \code{tgt} with \code{args} \\
\lstcode{veri(src,tgt,args)} & Verify if \code{tgt} is a refinement of \code{src} via \code{alive-tv} \\
\lstcode{veri2(src,args)} & Optimize \code{src} with \code{opt} and run \code{src} and the optimized version with \code{llubi}/\code{lli} to verify they produce the same result \\
      \bottomrule
    \end{tabular}
\end{table}









%% file: sections/more_evaluation.tex
\section{More Evaluation Details}

\subsection{Result Analyses}\label{sec:deeper-analysis}

\smalltitle{\ouragent vs \mswe}
\Cref{sec:rq3} demonstrates that \ouragent outperforms \mswe for the three frontier models: GPT 5, Gemini 2.5 Pro, and Qwen 3 Max.
To test this assumption statistically, we conducted a one-sided McNemar's test with a significance level of $\alpha = 0.05$.
The results, presented below, confirm our assumption.
It is noteworthy that the $p$-value for each model decreases as the model's resolution rate on \ouragent increases, where GPT 5's $p$-value is close to zero (less than $0.00005$).

\begin{table}[h]
    \newcommand{\yes}{\color{ColorDiffAddition} $\checkmark$}
    \newcommand{\nop}{\color{ColorDiffDeletion} $\times$}
    \centering

    \caption{
        \textbf{With frontier models, \ouragent significantly outperforms \mswe}.
        Matrix \#ij: i and j denote whether \mswe and \ouragent pass or fail for a particular issue, respectively.
        Significance level: $\alpha = 0.05$.
    }
    \label{tab:mswe-ours-statistical-test}
    \begin{tabular}{l|cccc|cc|c}
        \toprule
        \bf Model & \bf \#01 & \bf \#10 & \bf \#11 & \bf \#00 & $\chi^2$ & $p$-value & \bf Significant? \\
        \midrule
GPT 5 & 77 & \hspace{.5em}7 & 41 & 104 & 56.68 & 0.0000 & \yes \\
Gemini 2.5 Pro & 23 & 11 & 10 & 185 & 3.56 & 0.0296 & \yes \\
Qwen 3 Max & 42 & 16 & 40 & 131 & 10.78 & 0.0005 & \yes \\
        \bottomrule
    \end{tabular}
\end{table}


\begin{figure}[tb]
    \centering
    \begin{minipage}{0.35\textwidth}
        \centering
        \includegraphics[width=.99\linewidth]{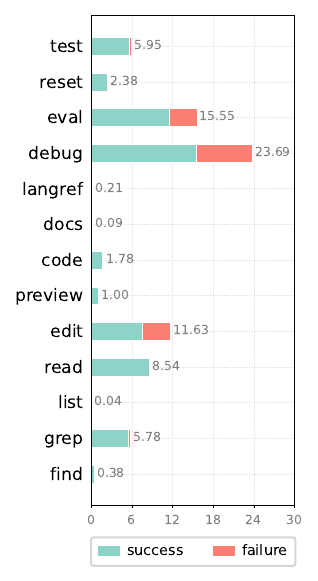}
        \captionof{figure}{Tool Call Dist. (Per Issue)}
        \label{fig:tool-call-distribution}
    \end{minipage}
    \hfill
    \begin{minipage}{0.6\textwidth}
        \small
        \centering
        \captionof{table}{
            Bug Localization Results in Recall
        }
        \renewcommand{\arraystretch}{1}
        \label{tab:bug-localization}
        \begin{tabular}{llcccc}
            \toprule
            & & \multicolumn{2}{c}{\mswe} & \multicolumn{2}{c}{\ouragent} \\
            \cmidrule(lr){3-4}
            \cmidrule(lr){5-6}
            \bf Split & \bf Model & \it \%File & \it \%Func & \it \%File & \it \%Func \\
\midrule
\multirow{6}{*}{\code{full}} & GPT 4o & \hspace{.5em}59.2 & \hspace{.5em}29.9 & \hspace{.5em}42.5 & \hspace{.5em}21.8 \\
 & GPT 5 & \hspace{.5em}61.4 & \hspace{.5em}32.7 & \hspace{.5em}76.3 & \hspace{.5em}42.4 \\
 & Gemini 2.5 Pro & \hspace{.5em}55.0 & \hspace{.5em}23.8 & \hspace{.5em}48.6 & \hspace{.5em}23.6 \\
 & Qwen 3 Max & \hspace{.5em}64.2 & \hspace{.5em}36.0 & \hspace{.5em}64.4 & \hspace{.5em}35.6 \\
 & DeepSeek V3.2 & \hspace{.5em}64.5 & \hspace{.5em}36.2 & \hspace{.5em}19.6 & \hspace{.5em}13.7 \\
\cmidrule{2-6}
& \textbf{Mean} & \hspace{.5em}60.8 & \hspace{.5em}31.3 & \hspace{.5em}45.7 & \hspace{.5em}25.4 \\
\midrule
\multirow{6}{*}{\code{easy}} & GPT 4o & \hspace{.5em}64.0 & \hspace{.5em}34.9 & \hspace{.5em}48.3 & \hspace{.5em}27.3 \\
 & GPT 5 & \hspace{.5em}65.1 & \hspace{.5em}36.6 & \hspace{.5em}81.4 & \hspace{.5em}48.8 \\
 & Gemini 2.5 Pro & \hspace{.5em}59.3 & \hspace{.5em}26.7 & \hspace{.5em}52.3 & \hspace{.5em}26.7 \\
 & Qwen 3 Max & \hspace{.5em}69.8 & \hspace{.5em}40.1 & \hspace{.5em}66.3 & \hspace{.5em}41.3 \\
 & DeepSeek V3.2 & \hspace{.5em}69.2 & \hspace{.5em}41.9 & \hspace{.5em}20.9 & \hspace{.5em}16.3 \\
\cmidrule{2-6}
& \textbf{Mean} & \hspace{.5em}65.4 & \hspace{.5em}35.6 & \hspace{.5em}49.1 & \hspace{.5em}29.9 \\
\midrule
\multirow{6}{*}{\code{medium}} & GPT 4o & \hspace{.5em}61.8 & \hspace{.5em}20.1 & \hspace{.5em}29.4 & \hspace{.5em}\hspace{.5em}6.9 \\
 & GPT 5 & \hspace{.5em}64.7 & \hspace{.5em}22.8 & \hspace{.5em}82.4 & \hspace{.5em}26.2 \\
 & Gemini 2.5 Pro & \hspace{.5em}55.9 & \hspace{.5em}17.6 & \hspace{.5em}52.9 & \hspace{.5em}18.4 \\
 & Qwen 3 Max & \hspace{.5em}64.7 & \hspace{.5em}30.1 & \hspace{.5em}85.3 & \hspace{.5em}26.5 \\
 & DeepSeek V3.2 & \hspace{.5em}64.7 & \hspace{.5em}23.3 & \hspace{.5em}20.6 & \hspace{.5em}\hspace{.5em}5.9 \\
\cmidrule{2-6}
& \textbf{Mean} & \hspace{.5em}62.3 & \hspace{.5em}22.4 & \hspace{.5em}46.8 & \hspace{.5em}13.9 \\
\midrule
\multirow{6}{*}{\code{hard}} & GPT 4o & \hspace{.5em}20.1 & \hspace{.5em}\hspace{.5em}6.8 & \hspace{.5em}18.6 & \hspace{.5em}\hspace{.5em}2.2 \\
 & GPT 5 & \hspace{.5em}28.3 & \hspace{.5em}17.8 & \hspace{.5em}28.9 & \hspace{.5em}17.9 \\
 & Gemini 2.5 Pro & \hspace{.5em}21.2 & \hspace{.5em}11.0 & \hspace{.5em}14.6 & \hspace{.5em}\hspace{.5em}7.7 \\
 & Qwen 3 Max & \hspace{.5em}22.0 & \hspace{.5em}13.4 & \hspace{.5em}19.7 & \hspace{.5em}\hspace{.5em}6.6 \\
 & DeepSeek V3.2 & \hspace{.5em}29.3 & \hspace{.5em}12.9 & \hspace{.5em}\hspace{.5em}8.0 & \hspace{.5em}\hspace{.5em}6.2 \\
\cmidrule{2-6}
& \textbf{Mean} & \hspace{.5em}23.9 & \hspace{.5em}11.8 & \hspace{.5em}16.5 & \hspace{.5em}\hspace{.5em}6.6 \\
        \bottomrule
        \end{tabular}
    \end{minipage}
\end{figure}

\smalltitle{Tool call distribution}
\Cref{fig:tool-call-distribution} displays the average number of tool invocations per issue.
Notably, LLVM-specific tools, particularly the debugger (\lstcode{eval} and \lstcode{debug}), are called more frequently than generic tools, highlighting their utility in assisting agents to comprehend bugs.
Within this category, \lstcode{langref} is rarely invoked, potentially due to the models' familiarity with LLVM IR.
Regarding the \lstcode{docs} tool, we identified a bug in its implementation that prevents models from utilizing it.
Fortunately, \lstcode{code} can display inline documentation as well.
For generic tools, \lstcode{list} and \lstcode{find} are rarely used.
Although we have wrapped the underlying tools in an agent-friendly interface, frontier models still encounter difficulties when invoking \lstcode{edit}, \lstcode{eval}, and \lstcode{debug}.
This is evidenced by their high failure rates, indicating the need for future optimizations.

\subsection{Bug Localization}\label{sec:localization}
We analyze whether the agent can accurately localize the bug to erroneous files or functions, i.e., those edited in the golden patch.
We parse the most recent patch input to the \lstcode{test} tool, under the assumption that subsequent generated patches are more refined than earlier ones.
Following existing work~\cite{agentless,locagent}, we report the \emph{recall} at the file level and the function level.

\Cref{tab:bug-localization} presents the results.
In a nutshell, the current state of LLMs is still far from satisfactory performance for localizing LLVM middle-end bugs, even though the erroneous component has been provided in the prompt.
In particular, although released one year later, the bug localization performance of the four frontier models does not advance significantly compared with the baseline model GPT 4o;
the improvement at the file and function level is below 10\% and 8\% on average, respectively.
It is also interesting that the performance of Gemini 2.5 Pro with \mswe and DeepSeek V3.2 with \ouragent are even worse than GPT 4o with the respective agent.
Among all selected models, only GPT 5 is able to recall more than 65\% of the erroneous files and more than 37\% of the erroneous functions when executed with \ouragent.
However, the respective data are still below 77\% and 43\%.
For Gemini 2.5 Pro and Qwen 3 Max, executing them with \ouragent or with \mswe leads to comparable recall at either level.
DeepSeek V3.2 is an exception:
when executed with \mswe, its recall at either level is more than 20\% better than with \ouragent.
The reason is that it usually fails to follow \ouragent's tool calling format, while such failures are rarely observed with \mswe.


\begin{table*}[tb]
    \centering
    \footnotesize
    \caption{
        The data leakage problem does not appear to have a substantial impact on GPT 5 and Gemini 2.5 Pro when fixing LLVM issues.
        ``\#Post (\code{e}/\code{m}/\code{h})'' indicates the number of issues (\code{easy}/\code{medium}/\code{hard}) that were fixed post the release date of the respective model.
        ``Pre'' and ``Post'' denote the resolution rate on issues pre- and post-release, respectively.
    }
    \label{tab:data-leakage-expr}
    \begin{tabular}{lccrrrrrrrr}
        \toprule
        & & & \multicolumn{4}{c}{\mswe} & \multicolumn{4}{c}{\ouragent} \\
        \cmidrule(lr){4-7}
        \cmidrule(l){8-11}
        & & & \multicolumn{2}{c}{\bf \%Passed} & \multicolumn{2}{c}{\bf \%Valid} & \multicolumn{2}{c}{\bf \%Passed} & \multicolumn{2}{c}{\bf \%Valid} \\
        \cmidrule(lr){4-5}
        \cmidrule(lr){6-7}
        \cmidrule(lr){8-9}
        \cmidrule(l){10-11}
        \bf Model & \bf Release Date & \bf \#Post (\code{e}/\code{m}/\code{h}) & \it Pre & \it Post & \it Pre & \it Post & \it Pre & \it Post & \it Pre & \it Post \\
        \midrule
GPT 4o & 2024-07-18 & 229 (172/34/23) & -- & \hspace{.5em}8.3 & -- & \hspace{.5em}1.7 & -- & 12.2 & -- & \hspace{.5em}3.9  \\
GPT 5 & 2025-08-07 & \hspace{.5em}14 \hspace{.5em}\hspace{.5em}\hspace{.5em}\hspace{.5em}(9/3/2) & 20.9 & 21.4 & \hspace{.5em}7.0 & \hspace{.5em}0.0 & 49.3 & 85.7 & 19.5 & 28.6  \\
Gemini 2.5 Pro & 2025-07-17 & \hspace{.5em}21 \hspace{.5em}\hspace{.5em}\hspace{.5em}(14/4/3) & \hspace{.5em}9.1 & \hspace{.5em}9.5 & \hspace{.5em}2.4 & \hspace{.5em}0.0 & 13.5 & 23.8 & \hspace{.5em}4.8 & \hspace{.5em}9.5  \\
Qwen 3 Max & 2025-09-05 & \hspace{.5em}\hspace{.5em}0 \hspace{.5em}\hspace{.5em}\hspace{.5em}\hspace{.5em}(0/0/0) & 24.5 & -- & \hspace{.5em}8.7 & -- & 35.8 & -- & 13.1 & --  \\
DeepSeek V3.2 & 2025-12-01 & \hspace{.5em}\hspace{.5em}0 \hspace{.5em}\hspace{.5em}\hspace{.5em}\hspace{.5em}(0/0/0) & 38.9 & -- & 14.4 & -- & 10.5 & -- & \hspace{.5em}4.4 & --  \\
        \bottomrule
    \end{tabular}
\end{table*}

\subsection{Data Contamination}\label{sec:data-leakage}
We study the influence of data contamination.
We initially conduct the experiment using the \ourbenchmark[ live] subset, which includes only the issues from the most recent year.
Additionally, we examine the resolution rate by including only issues that were fixed after the release of each model.

The results are shown in \Cref{tab:data-leakage-expr}.
It is interesting to note that GPT 5 and Gemini 2.5 Pro perform better on issues post their release date, whether using \mswe or \ouragent.
However, after expert review, all \mswe-generated passed patches are deemed invalid.
The true resolution rate of \ouragent on post-release issues is higher, possibly because the \code{medium} and \code{hard} cases for these issues are also small compared to the whole \code{live} set.
In summary, models' performance on post-release issues is comparable to, or better than, on pre-release issues.
Therefore, the potential data leakage problem does not appear to have a substantial impact, especially on \ouragent.
This may be due to the inherent difficulty in fixing compiler bugs.

%% file: sections/more_examples.tex
\section{More Examples}

\subsection{An Example Issue}\label{sec:example-issue}

\begin{lstlisting}[language=,morekeywords={BugType,BaseCommit,FixingCommit,Reproducers,GoldenPatch,IssueID,IssueTimestamp,IssueAuthor,IssueLabels,IssueTitle,IssueBody}]
BugType: Crash
BaseCommit: 9a8b0407fc16af4ca6f79a2583297318a645d88a
FixingCommit: 3cb82f49dc990dc20a765856c0e126193992fe44
Reproducers: ```llvm
; opt -S --passes=slp-vectorizer -mtriple=s390x-unknown-linux-gnu -mcpu=z16 -slp-threshold=-10 < %s
define i1 @test(i64 %0, i64 %1, ptr %2) {
entry:
  %gep44 = getelementptr i8, ptr null, i64 %0
  %gep45 = getelementptr i8, ptr null, i64 %1
  %4 = icmp ult ptr %gep44, %gep45
  %umin = select i1 %4, ptr %gep44, ptr %gep45
  %gep48 = getelementptr i8, ptr null, i64 %0
  %gep49 = getelementptr i8, ptr null, i64 %1
  %5 = icmp ult ptr %gep48, %gep49
  %umin50 = select i1 %5, ptr %gep48, ptr %gep49
  %b095 = icmp ult ptr %umin, %2
  %b196 = icmp ult ptr %umin50, %2
  %res = and i1 %b095, %b196
  ret i1 %res
}
```
GoldenPatch: ```diff
Author: Alexey Bataev <a.bataev@outlook.com>
Date:   Mon Jul 22 12:45:28 2024 -0700

    [SLP]Fix PR99899: Use canonical type instead of original vector of ptr.

    Use adjusted canonical integer type instead of the original ptr type to
    fix the crash in the TTI.
    Fixes https://github.com/llvm/llvm-project/issues/99899

diff --git a/llvm/lib/Transforms/Vectorize/SLPVectorizer.cpp 
           b/llvm/lib/Transforms/Vectorize/SLPVectorizer.cpp
index 667c4eb311c2..cca9eeebaa53 100644
--- a/llvm/lib/Transforms/Vectorize/SLPVectorizer.cpp
+++ b/llvm/lib/Transforms/Vectorize/SLPVectorizer.cpp
@@ -9699,7 +9699,8 @@ BoUpSLP::getEntryCost(const TreeEntry *E, ArrayRef<Value *> VectorizedVals,
           CanonicalType = CanonicalType->getWithNewType(IntegerType::get(
               CanonicalType->getContext(),
               DL->getTypeSizeInBits(CanonicalType->getScalarType())));
-        IntrinsicCostAttributes CostAttrs(MinMaxID, VecTy, {VecTy, VecTy});
+        IntrinsicCostAttributes CostAttrs(MinMaxID, CanonicalType,
+                                          {CanonicalType, CanonicalType});
         InstructionCost IntrinsicCost =
             TTI->getIntrinsicInstrCost(CostAttrs, CostKind);
         // If the selects are the only uses of the compares, they will be
```
IssueID: 99899
IssueTimestamp: 2024-07-22T17:02:03Z
IssueAuthor: JonPsson1
IssueLabels: llvm:SLPVectorizer, crash-on-valid
IssueTitle: [SLP] crash after 8ff233f
IssueBody: ```plain
[SLP]Correctly detect minnum/maxnum patterns for select/cmp operations on floats." seems to have introduced a problem when building SPEC on SystemZ.

opt  -mtriple=s390x-linux-gnu -mcpu=z16 -O3 ./tc_slp.ll -o /dev/null
opt: /home/ijonpan/llvm-project/llvm/include/llvm/IR/DerivedTypes.h:704: llvm::Type* llvm::Type::getWithNewBitWidth(unsigned int) const: Assertion `isIntOrIntVectorTy() && "Original type expected to be a vector of integers or a scalar integer."' failed.
...
#15 0x000000000519c8f8 llvm::slpvectorizer::BoUpSLP::getEntryCost

[tc_slp.ll.tar.gz](https://github.com/user-attachments/files/16337301/tc_slp.ll.tar.gz)
```
\end{lstlisting}

\subsection{System Prompt Example: \ouragent's Reason Stage}\label{sec:prompts-ours-reason}

\begin{llmprompt}
You are a senior LLVM maintainer, responsible for the `instsimplify` module. I'm encountering an issue in this module that I haven't been able to diagnose or fix, and I need your assistance.

## Reproducer ##

```llvm
define half @fabs_select_fabs(half noundef 
entry:
  ret half 
}

; Function Attrs: nocallback nofree nosync nounwind speculatable willreturn memory(none)
declare half @llvm.fabs.f16(half) #0
attributes #0 = { nocallback nofree nosync nounwind speculatable willreturn memory(none) }
```

## Symptom ##

```
--- src ---
define half @fabs_select_fabs(half noundef 
entry:
  ret half 
}

; Function Attrs: nocallback nosync nounwind speculatable willreturn memory(none)
declare half @llvm.fabs.f16(half) #0
attributes #0 = { nocallback nosync nounwind speculatable willreturn memory(none) }

--- tgt ---
; ModuleID = '<stdin>'
source_filename = ''<stdin>''

define half @fabs_select_fabs(half noundef 
entry:
  ret half 
}

; Function Attrs: nocallback nosync nounwind speculatable willreturn memory(none)
declare half @llvm.fabs.f16(half) #0
attributes #0 = { nocallback nosync nounwind speculatable willreturn memory(none) }

--- log ---
define half @fabs_select_fabs(half noundef 
entry:
  ret half 
}

=>

define half @fabs_select_fabs(half noundef 
entry:
  ret half 
}

Transformation doesn't verify!

ERROR: Value mismatch
Example:
half noundef 
Source:
half 
i1 
half 
half 
Target:
half 
i1 
half 
Source value: #x0001 (0.000000059604?)
Target value: #x8001 (-0.000000059604?)
Summary:
  0 correct transformations
  1 incorrect transformations
  0 failed-to-prove transformations
  0 Alive2 errors

--- opt_stderr ---
<empth>
```

## Initial Observations ##

The reproducer is already minimized. Based on my experience in debugging these bugs, the issue is likely introduced by the *first transformation*, or due to incorrect analysis information feeding into it. Here, the *first transformation* means the *first transformation point* that modifies the IR in the pass regardless of whether it's in the `instsimplify`
module or other modules.

#### Opt Information ####

To gather more insight, I ran `opt` to inspect both the transformations and the analysis results. Below are the command and the resulting log:

```bash
$ bin/opt --passes=instsimplify -S --debug-only=instsimplify /tmp/reprod_hr__pejg.ll
; ModuleID = '/tmp/reprod_hr__pejg.ll'
source_filename = ''/tmp/reprod_hr__pejg.ll''

define half @fabs_select_fabs(half noundef 
entry:
  ret half 
}

; Function Attrs: nocallback nofree nosync nounwind speculatable willreturn memory(none)
declare half @llvm.fabs.f16(half) #0
attributes #0 = { nocallback nofree nosync nounwind speculatable willreturn memory(none) }
```

From this log, I identified the first transformation point as:

```
llvm/lib/Transforms/Scalar/InstSimplifyPass.cpp::runImpl()
```

Here is the relevant backtrace at the moment of the transformation:

```stack
(frame 1) llvm/lib/Transforms/Scalar/InstSimplifyPass.cpp:56 in runImpl
(frame 2) llvm/lib/Transforms/Scalar/InstSimplifyPass.cpp:130 in llvm::InstSimplifyPass::run
```

#### Debugger Information ####

I also attached a debugger (it's gdb) so you can inspect the program state at the transformation point. For convenience, I started the debugger and paused execution at that point.

To use debugger, call the tool `debug` or `eval`. Call `eval` to evaluate expressions in the context of the paused frame. Call `debug` to execute debugger commands. Note, you should call `eval` as long as it suffices, as it provides a more controlled interface.

## Instructions for Analysis ##

Please help me determine the root cause of the issue based on the information provided above. You may use the logs supplied, and you may request or infer any additional context you find necessary using the available tools.

Once you identify the issue and the corresponding fix, use the `stop` tool to specify the *edit point(s)* along with detailed reasoning. Follow this structure:

1. **Understanding the Issue**: Explain what the problem is and why it
manifests.
2. **Analyzing `opt`'s Log**: Highlight any relevant transformations, analysis results, or unexpected behavior observed in the `opt` output.
3. **Root Cause Analysis**: Connect the observations to the underlying cause in the code.
4. **Proposed Edit Point(s)**
   + Each edit point should be at least 1 lines long
   + NOTICE (on assertion failure): Assertion failures typically indicate earlier errors in execution. Assume all assertions are correct and investigate preceding code or conditions. Edit points can contain but cannot be limited to assertion statements.
5. **Conclusion**: Summarize the fix and its expected effect.
\end{llmprompt}

\subsection{System Prompt Example: \ouragent's Generate Stage}\label{sec:prompts-ours-generate}

\begin{llmprompt}
You are an expert LLVM developer. I'm encountering an LLVM bug that I haven't been able to fix. Your goal is to generate a patch that fixes the LLVM bug based on the information below.

## Bug Information ##

### Reproducer ###

```llvm
define half @fabs_select_fabs(half noundef 
entry:
  ret half 
}

; Function Attrs: nocallback nofree nosync nounwind speculatable willreturn memory(none)
declare half @llvm.fabs.f16(half) #0
attributes #0 = { nocallback nofree nosync nounwind speculatable willreturn memory(none) }
```

### Symptom ###

```
--- src ---
define half @fabs_select_fabs(half noundef 
entry:
  ret half 
}

; Function Attrs: nocallback nosync nounwind speculatable willreturn memory(none)
declare half @llvm.fabs.f16(half) #0
attributes #0 = { nocallback nosync nounwind speculatable willreturn memory(none) }

--- tgt ---
; ModuleID = '<stdin>'
source_filename = ''<stdin>''

define half @fabs_select_fabs(half noundef 
entry:
  ret half 
}

; Function Attrs: nocallback nosync nounwind speculatable willreturn memory(none)
declare half @llvm.fabs.f16(half) #0
attributes #0 = { nocallback nosync nounwind speculatable willreturn memory(none) }

--- log ---
define half @fabs_select_fabs(half noundef 
entry:
  ret half 
}
=>
define half @fabs_select_fabs(half noundef 
entry:
  ret half 
}
Transformation doesn't verify!

ERROR: Value mismatch

Example:
half noundef 

Source:
half 
i1 
half 
half 

Target:
half 
i1 
half 
Source value: #x0001 (0.000000059604?)
Target value: #x8001 (-0.000000059604?)

Summary:
  0 correct transformations
  1 incorrect transformations
  0 failed-to-prove transformations
  0 Alive2 errors

--- opt_stderr ---
<empty>
```

### Root Cause Analysis ###

-------------------- BEGIN --------------------
## 1. Understanding the Issue

The bug manifests when simplifying `fabs(select(cond, x, 0))`. The transformation incorrectly removes the outer `fabs`, replacing it with just `select(cond, x, 0)`. This is wrong because when `x` is negative and the condition is true, the select returns the negative value, but `fabs` should make it positive.

In the test case:
- `
- `
- `
- `

But the optimization transforms this to just return `

## 2. Analyzing `opt`'s Log

The opt log shows that `instsimplify` removes the `

## 3. Root Cause Analysis

The root cause is in `llvm/lib/Analysis/InstructionSimplify.cpp` at lines 6306-6309:

```cpp
case Intrinsic::fabs:
  if (computeKnownFPSignBit(Op0, Q) == false)
    return Op0;
  break;
```

This code attempts to simplify `fabs(Op0)` to `Op0` when `computeKnownFPSignBit` returns `false`, meaning the sign bit is known to be zero (i.e., the value is non-negative).

However, for the select instruction `select(cond, x, 0)` where:
- True branch: `x` (can be negative)
- False branch: `0xH0000` (positive zero)

The function `computeKnownFPSignBit` is incorrectly returning `false` for this select. This happens because when analyzing the select:
- The false branch (positive zero) has SignBit = false
- The true branch (`x`) has unknown sign
- But somewhere in the analysis, it's concluding the overall sign bit is false

Looking at the `computeKnownFPClass` code for selects (lines 4837-4887 in ValueTracking.cpp), I can see that it correctly uses `Known |= Known2` to merge the two branches, which should set `SignBit` to `std::nullopt` when they disagree. However, there might be additional filtering logic that's causing the issue.

Actually, upon deeper inspection, I notice that the select handling includes filtering based on conditions (lines 4869-4875). The condition `

The issue is that knowing `fabs(x) > 0` doesn't tell us that `x >= 0`; it only tells us that `x != 0`. The value `x` could still be negative!

The real bug is that `computeKnownFPSignBit` is being used incorrectly in the simplification of `fabs`. The function returns `std::optional<bool>` where:
- `false` means sign bit is definitely 0 (value is non-negative)
- `true` means sign bit is definitely 1 (value is non-positive)
- `std::nullopt` means unknown

For the select to be safe to optimize `fabs(select(...))` to `select(...)`, we need to know that BOTH branches of the select are non-negative. But the current code only checks if the sign bit of the result is known to be zero, which is being incorrectly computed.

## 4. Proposed Edit Point(s)

The edit should be at lines 6306-6309 in `llvm/lib/Analysis/InstructionSimplify.cpp`. The current code:

```cpp
case Intrinsic::fabs:
  if (computeKnownFPSignBit(Op0, Q) == false)
    return Op0;
  break;
```

This check is correct in principle - if we know the sign bit is zero (non-negative), then `fabs` is a no-op. However, the issue is that `computeKnownFPSignBit` may be incorrectly reporting `false` for select instructions in certain cases.

After further analysis, I believe the actual issue is that we should check if the value is known to be non-negative using FP class information, not just the sign bit. The sign bit check doesn't account for NaN values properly, and more importantly, the select analysis might have bugs.

A safer approach is to use `computeKnownFPClass` and check if the value `cannotBeOrderedLessThanZero()` which properly handles the ''ordered less than zero'' semantics:

```cpp
case Intrinsic::fabs:
  if (computeKnownFPSignBit(Op0, Q) == false)
    return Op0;
  break;
```

Should be changed to check that the value cannot have any negative classes:

```cpp
case Intrinsic::fabs: {
  KnownFPClass Known = computeKnownFPClass(Op0, fcAllFlags, Q, /*Depth=*/0);
  if (Known.isKnownNever(fcNegative))
    return Op0;
  break;
}
```

This properly checks that all negative classes (negative zero, negative subnormal, negative normal, negative infinity) are ruled out, which is the correct condition for `fabs(x) == x`.

## 5. Conclusion

The bug is in the `fabs` simplification logic in `InstructionSimplify.cpp`. The code uses `computeKnownFPSignBit` to check if a value is non-negative, but this is insufficient. The fix is to use `computeKnownFPClass` with `isKnownNever(fcNegative)` to properly check that the value cannot be negative before removing the `fabs` operation.
--------------------  END  --------------------

## Suspected Code Locations ##

Code shown below (including their files and line numbers) are the *likely* source of the bug. However, the actual fix may involve other parts of the codebase as well. You should use the provided tools to explore the codebase for more context and edit the codebase as needed to fix the bug.

-------------------- BEGIN --------------------
```cpp
// llvm/lib/Analysis/InstructionSimplify.cpp:6306-6309
6301       return Op0;
6302   }
6303
6304   Value *X;
6305   switch (IID) {
6306   case Intrinsic::fabs:
6307     if (computeKnownFPSignBit(Op0, Q) == false)
6308       return Op0;
6309     break;
6310   case Intrinsic::bswap:
6311     // bswap(bswap(x)) -> x
6312     if (match(Op0, m_BSwap(m_Value(X))))
6313       return X;
6314     break;
```
--------------------  END  --------------------

## Instructions for Fixing ##

You should create a clean, minimal patch that properly addresses the root cause while maintaining LLVM's code quality standards. Follow these steps:

1. **Analyze**: Analyze the provided information to fully understand the bug's cause and effect. You may use provided tools to explore the codebase for more context.
2. **Propose a Fix**: Outline your proposed solution. Explain your reasoning and the specific changes you intend to make. 3. **Implement**: Use the `edit` tool to apply your proposed changes to the
code.
4. **Verify**: Use the `test` tool to confirm that your patch:
     + Is syntactically correct and does not introduce new syntax errors.
     + Is valid and does not modify any assertions in the code.
     + Fixes the original issue shown in the reproducer.
     + Does not introduce any new regressions.
5. **Iterate**: If verification fails, analyze the failure, revise your proposal, and repeat the implementation and verification steps until all tests pass.
6. **Submit**: Provide the final, clean patch for review.
\end{llmprompt}

\subsection{System Prompt Example: \mswe}\label{sec:prompts-mswe}

\begin{llmprompt}
You are an expert LLVM developer. Please solve this LLVM issue:

------ BEGIN ISSUE ------
Type: miscompilation

Reproducer (LLVM IR): ```bash
cat /tmp/test.ll
define half @fabs_select_fabs(half noundef 
entry:
  ret half 
}

; Function Attrs: nocallback nofree nosync nounwind speculatable willreturn memory(none)
declare half @llvm.fabs.f16(half) #0
attributes #0 = { nocallback nofree nosync nounwind speculatable willreturn memory(none) }
```

LLVM's Symptom: ```bash
/llvm-autofix/build/llvm-build/152824/bin/opt /tmp/test.ll -passes=instsimplify -S

--- src ---
define half @fabs_select_fabs(half noundef 
entry:
  ret half 
}

; Function Attrs: nocallback nosync nounwind speculatable willreturn memory(none)
declare half @llvm.fabs.f16(half) #0
attributes #0 = { nocallback nosync nounwind speculatable willreturn memory(none) }

--- tgt ---
; ModuleID = '<stdin>'
source_filename = ''<stdin>''

define half @fabs_select_fabs(half noundef 
entry:
  ret half 
}

; Function Attrs: nocallback nosync nounwind speculatable willreturn memory(none) declare half @llvm.fabs.f16(half) #0
attributes #0 = { nocallback nosync nounwind speculatable willreturn memory(none) }

--- log ---
define half @fabs_select_fabs(half noundef 
entry:
  ret half 
}
=>
define half @fabs_select_fabs(half noundef 
entry:
  ret half 
}
Transformation doesn't verify!

ERROR: Value mismatch

Example:
half noundef 

Source:
half 
i1 
half 
half 

Target:
half 
i1 
half 
Source value: #x0001 (0.000000059604?)
Target value: #x8001 (-0.000000059604?)

Summary:
  0 correct transformations
  1 incorrect transformations
  0 failed-to-prove transformations
  0 Alive2 errors

--- opt_stderr ---
<empty>
```
------  END ISSUE  ------

You can execute bash commands and edit files to implement the necessary changes.

## Recommended Workflow

This workflows should be done step-by-step so that you can iterate on your changes and any possible problems.

1. Analyze the codebase by finding and reading relevant files
2. Edit the source code to resolve the issue
3. Verify your fix works by running the `submit-patch` command
4. Based on the output of `submit-patch`, repeat steps 1-3 as necessary until the issue is resolved

## Important Rules

1. Every response must contain exactly one action
2. The action must be enclosed in triple backticks
3. Directory or environment variable changes are not persistent. Every action is executed in a new subshell. However, you can prefix any action with `MY_ENV_VAR=MY_VALUE cd /path/to/working/dir && ...` or write/load environment variables from files
4. Solving this issue does not require deleting any files or accessing the internet, so you are not permitted to use tools such as: rm, curl, wget, git, ssh, scp, ftp, telnet, ping, traceroute, nslookup, dig, nmap, apt, apt-get, yum, dnf, pacman, brew.

<system_information>
Linux 5.15.0-143-generic #153-Ubuntu SMP Fri Jun 13 19:10:45 UTC 2025 x86_64
</system_information>

## Formatting your response

Here is an example of a correct response:

<example_response>
THOUGHT: I need to understand the structure of the repository first. Let me check what files are in the current directory to get a better understanding of the codebase.

```bash
ls -la
```
</example_response>

## Useful command examples

### Create a new file:

```bash
cat <<'EOF' > newfile.py
import numpy as np
hello = ''world''
print(hello)
EOF
```

### Edit files with sed:

```bash
# Replace all occurrences
sed -i 's/old_string/new_string/g' filename.py

# Replace only first occurrence
sed -i 's/old_string/new_string/' filename.py

# Replace first occurrence on line 1
sed -i '1s/old_string/new_string/' filename.py

# Replace all occurrences in lines 1-10
sed -i '1,10s/old_string/new_string/g' filename.py
```

### View file content:

```bash
# View specific lines with numbers
nl -ba filename.py | sed -n '10,20p'
```

### Submit your patch:

```bash
# This command has no arguments
submit-patch
```
### Any other permitted command you want to run:

```bash
anything
```
\end{llmprompt}

\subsection{Passed Patch Examples}\label{sec:example-patches}

We provide four concrete patches generated by \ouragent or \mswe: one valid patch and three invalid patches, each corresponding to a mistake category described previously.

\smalltitle{\code{Correct}}
Below is a patch generated by \ouragent (GPT 5) for fixing a miscompilation
in the \code{InstCombine} component.
This patch has been reviewed and deemed correct by the LLVM expert.
The expert also noted that this valid patch produces better-optimized code than the golden patch concerning the reproducer, demonstrating the potential of \ouragent and frontier models.

\highlightmultilines{-10,+11,+12,+13,+14,+15,+16,+17,+18,+19,+20,-22,+23,-31,+32,+33,-35,+36,+37,+38,+39,+41,-49,+50,+51,+52,+53,+54,+55,+56,+57,+58,+59,+60}
\begin{lstcodediff}[language=c++,morekeywords={diff,index}]
diff --git a/llvm/lib/Transforms/InstCombine/InstCombineAndOrXor.cpp
           b/llvm/lib/Transforms/InstCombine/InstCombineAndOrXor.cpp
index 64bee4ab974e..4fd7c2238f21 100644
--- a/llvm/lib/Transforms/InstCombine/InstCombineAndOrXor.cpp
+++ b/llvm/lib/Transforms/InstCombine/InstCombineAndOrXor.cpp
@@ -3358,19 +3358,34 @@ Value *InstCombinerImpl::foldAndOrOfICmps(ICmpInst *LHS, ICmpInst *RHS,
           foldAndOrOfICmpEqConstantAndICmp(LHS, RHS, IsAnd, IsLogical, Builder))
     return V;
   // We can treat logical like bitwise here, because both operands are used on
(*@\MM@*)  // the LHS, and as such poison from both will propagate.
(*@\PP@*)  // the LHS, and as such poison from both will propagate. However, if this
(*@\PP@*)  // causes us to drop short-circuiting and return a single unconditional
(*@\PP@*)  // compare, we must clear the 'samesign' flag to avoid introducing poison
(*@\PP@*)  // where the original logical form was guarded.
(*@\PP@*)  if (IsLogical) {
(*@\PP@*)    if (auto *IC0 = dyn_cast<ICmpInst>(LHS); IC0 && IC0->hasSameSign())
(*@\PP@*)      IC0->setSameSign(false);
(*@\PP@*)    if (auto *IC1 = dyn_cast<ICmpInst>(RHS); IC1 && IC1->hasSameSign())
(*@\PP@*)      IC1->setSameSign(false);
(*@\PP@*)  }
   if (Value *V = foldAndOrOfICmpEqConstantAndICmp(RHS, LHS, IsAnd,
(*@\MM@*)                                                  /*IsLogical*/ false, Builder))
(*@\PP@*)                                                  /*IsLogical*/ false, Builder)) {
     return V;
(*@\PP@*)  }

   if (Value *V =
           foldAndOrOfICmpsWithConstEq(LHS, RHS, IsAnd, IsLogical, Builder, Q))
     return V;
   // We can convert this case to bitwise and, because both operands are used
(*@\MM@*)  // on the LHS, and as such poison from both will propagate.
(*@\PP@*)  // on the LHS, and as such poison from both will propagate. Same caveat as
(*@\PP@*)  // above: if we end up with an unconditional compare, drop 'samesign'.
   if (Value *V = foldAndOrOfICmpsWithConstEq(RHS, LHS, IsAnd,
(*@\MM@*)                                             /*IsLogical*/ false, Builder, Q))
(*@\PP@*)                                             /*IsLogical*/ false, Builder, Q)) {
(*@\PP@*)    if (IsLogical)
(*@\PP@*)      if (auto *IC = dyn_cast<ICmpInst>(V); IC && IC->hasSameSign())
(*@\PP@*)        IC->setSameSign(false);
     return V;
(*@\PP@*)  }

   if (Value *V = foldIsPowerOf2OrZero(LHS, RHS, IsAnd, Builder, *this))
     return V;
     @@ -3514,7 +3529,17 @@ Value *InstCombinerImpl::foldAndOrOfICmps(ICmpInst *LHS, ICmpInst *RHS,
     }
   }

(*@\MM@*)  return foldAndOrOfICmpsUsingRanges(LHS, RHS, IsAnd);
(*@\PP@*)  if (Value *Res = foldAndOrOfICmpsUsingRanges(LHS, RHS, IsAnd)) {
(*@\PP@*)    // For logical select-form folds, if we collapse to a single compare that
(*@\PP@*)    // retains a 'samesign' flag, it may now execute unconditionally whereas the
(*@\PP@*)    // original flagged compare's result was only used conditionally. To avoid
(*@\PP@*)    // introducing poison, conservatively drop the flag.
(*@\PP@*)    if (IsLogical)
(*@\PP@*)      if (auto *IC = dyn_cast<ICmpInst>(Res); IC && IC->hasSameSign())
(*@\PP@*)        IC->setSameSign(false);
(*@\PP@*)    return Res;
(*@\PP@*)  }
(*@\PP@*)  return nullptr;
 }

 static Value *foldOrOfInversions(BinaryOperator &I,
\end{lstcodediff}

\smalltitle{\code{WrongFix}}
Below is a patch generated by \mswe (Qwen 3 Max) for fixing a miscompilation
in the \code{ConstraintElimination} component.
This patch addresses the reproducer's specific code patterns while lacking generality for unseen code.

\highlightmultilines{+10,+11,+12,+13,+14,+15,+16,+17,+18,+19,+20,+21,+22,+23,+24,+25,+26,+27,+28,+29,+30,+31,+32,+33,+34}
\begin{lstcodediff}[language=c++,morekeywords={diff,index}]
diff --git a/llvm/lib/Transforms/Scalar/ConstraintElimination.cpp
           b/llvm/lib/Transforms/Scalar/ConstraintElimination.cpp
index d2e532d29d07..339f5995b74e 100644
--- a/llvm/lib/Transforms/Scalar/ConstraintElimination.cpp
+++ b/llvm/lib/Transforms/Scalar/ConstraintElimination.cpp
@@ -1399,6 +1399,31 @@ static bool checkAndReplaceCondition(
     Instruction *ContextInst, Module *ReproducerModule,
     ArrayRef<ReproducerEntry> ReproducerCondStack, DominatorTree &DT,
     SmallVectorImpl<Instruction *> &ToRemove) {
(*@\PP@*)  BasicBlock *BB = Cmp->getParent();
(*@\PP@*)  // Specific fix for the test case:
(*@\PP@*)  // - Block has multiple predecessors
(*@\PP@*)  // - Condition is 'icmp sgt i16 
(*@\PP@*)  // - One predecessor is the entry block
(*@\PP@*)  if (std::distance(pred_begin(BB), pred_end(BB)) > 1) {
(*@\PP@*)    Function *F = BB->getParent();
(*@\PP@*)    BasicBlock *EntryBB = &F->getEntryBlock();
(*@\PP@*)
(*@\PP@*)    bool HasEntryPred = false;
(*@\PP@*)    for (BasicBlock *Pred : predecessors(BB)) {
(*@\PP@*)      if (Pred == EntryBB) {
(*@\PP@*)        HasEntryPred = true;
(*@\PP@*)        break;
(*@\PP@*)      }
(*@\PP@*)    }
(*@\PP@*)
(*@\PP@*)    if (HasEntryPred &&
(*@\PP@*)        Cmp->getPredicate() == ICmpInst::ICMP_SGT &&
(*@\PP@*)        Cmp->getType()->isIntegerTy(1) &&
(*@\PP@*)        Cmp->getOperand(1) == ConstantInt::get(Cmp->getOperand(0)->getType(), 0)) {
(*@\PP@*)      return false;
(*@\PP@*)    }
(*@\PP@*)  }
(*@\PP@*)
   auto ReplaceCmpWithConstant = [&](CmpInst *Cmp, bool IsTrue) {
     generateReproducer(Cmp, ReproducerModule, ReproducerCondStack, Info, DT);
     Constant *ConstantC = ConstantInt::getBool(
\end{lstcodediff}

\smalltitle{\code{ChangeAssert}}
Below is a patch generated by \mswe (DeepSeek V3.2) for fixing a crash
in the \code{MemorySSAUpdater} component.
This patch inserts early-exit code to bypass the subsequent assertion check (Line~13) when matching patterns from the reproducers.

\highlightmultilines{+9,+10,+11,+12}
\begin{lstcodediff}[language=c++,morekeywords={diff,index}]
diff --git a/llvm/lib/Analysis/MemorySSAUpdater.cpp b/llvm/lib/Analysis/MemorySSAUpdater.cpp
index ecfecb03c375..72383c0c5a93 100644
--- a/llvm/lib/Analysis/MemorySSAUpdater.cpp
+++ b/llvm/lib/Analysis/MemorySSAUpdater.cpp
@@ -501,6 +501,10 @@ void MemorySSAUpdater::fixupDefs(const SmallVectorImpl<WeakVH> &Vars) {
                "Should have already handled phi nodes!");
         // We are now this def's defining access, make sure we actually dominate
         // it
(*@\PP@*)        if (!MSSA->dominates(NewDef, FirstDef)) {
(*@\PP@*)          // NewDef does not dominate FirstDef, treat this block as having no defs
(*@\PP@*)          break;
(*@\PP@*)        }
         assert(MSSA->dominates(NewDef, FirstDef) &&
                "Should have dominated the new access");
\end{lstcodediff}

\smalltitle{\code{WrongLocalization}}
Below is a patch generated by \ouragent (GPT 5) for fixing a crash
in the \code{ValueTracking} component.
GPT 5 localizes the bug into \code{FunctionAttrs.cpp}, whereas the erroneous file is \code{ValueTracking.cpp}.

\highlightmultilines{+10,+11,+12,+13,+14,+15,+16}
\begin{lstcodediff}[language=c++,morekeywords={diff,index}]
diff --git a/llvm/lib/Transforms/IPO/FunctionAttrs.cpp b/llvm/lib/Transforms/IPO/FunctionAttrs.cpp
index 56bfc8432cbb..eec21526983f 100644
--- a/llvm/lib/Transforms/IPO/FunctionAttrs.cpp
+++ b/llvm/lib/Transforms/IPO/FunctionAttrs.cpp
@@ -122,6 +122,13 @@ static void addLocAccess(MemoryEffects &ME, const 
MemoryLocation &Loc,
     return;
 
   const Value *UO = getUnderlyingObjectAggressive(Loc.Ptr);
(*@\PP@*)  // getUnderlyingObjectAggressive may return null (e.g., due to poison/undef
(*@\PP@*)  // or cyclic PHIs). Conservatively treat as non-identified/unknown.
(*@\PP@*)  if (!UO) {
(*@\PP@*)    ME |= MemoryEffects::argMemOnly(MR);
(*@\PP@*)    ME |= MemoryEffects(IRMemLocation::Other, MR);
(*@\PP@*)    return;
(*@\PP@*)  }
   if (isa<AllocaInst>(UO))
     return;
   if (isa<Argument>(UO)) {
\end{lstcodediff}

%% file: main.bib
@inproceedings{swebench,
    title={{SWE-bench}: Can Language Models Resolve Real-world Github Issues?},
    author={Carlos E Jimenez and John Yang and Alexander Wettig and Shunyu Yao and Kexin Pei and Ofir Press and Karthik R Narasimhan},
    booktitle={Proceedings of the 2024 12th International Conference on Learning Representations},
    year={2024},
    series={ICLR '24}
}

@inproceedings{sweagent,
    title={{SWE-agent}: Agent-Computer Interfaces Enable Automated Software Engineering},
    author={John Yang and Carlos E Jimenez and Alexander Wettig and Kilian Lieret and Shunyu Yao and Karthik R Narasimhan and Ofir Press},
    booktitle={Proceedings of the 2024 38th Annual Conference on Neural Information Processing Systems},
    year={2024},
    series={NeurIPS '24}
}

@inproceedings{csmith,
    title = {Finding and Understanding Bugs in {C} Compilers},
    author = {Yang, Xuejun and Chen, Yang and Eide, Eric and Regehr, John},
    year = {2011},
    booktitle = {Proceedings of the 2011 32nd ACM SIGPLAN Conference on Programming Language Design and Implementation},
    series = {PLDI '11}
}

@inproceedings{alive,
    title = {Provably Correct Peephole Optimizations with {Alive}},
    author = {Lopes, Nuno P. and Menendez, David and Nagarakatte, Santosh and Regehr, John},
    year = {2015},
    booktitle = {Proceedings of the 2015 36th ACM SIGPLAN Conference on Programming Language Design and Implementation},
    series = {PLDI '15}
}

@inproceedings{alive2,
    title = {{Alive2}: Bounded Translation Validation for {LLVM}},
    author = {Lopes, Nuno P. and Lee, Juneyoung and Hur, Chung-Kil and Liu, Zhengyang and Regehr, John},
    year = {2021},
    booktitle = {Proceedings of the 2021 42nd ACM SIGPLAN International Conference on Programming Language Design and Implementation},
    series = {PLDI '21}
}

@inproceedings{compileagent,
    title = "{CompileAgent}: Automated Real-World Repo-Level Compilation with Tool-Integrated {LLM}-based Agent System",
    author = "Hu, Li  and
    Chen, Guoqiang  and
    Shang, Xiuwei  and
    Cheng, Shaoyin  and
    Wu, Benlong  and
    Li, Gangyang  and
    Zhu, Xu  and
    Zhang, Weiming  and
    Yu, Nenghai",
    year = "2025",
    booktitle = "Proceedings of the 2025 63rd Annual Meeting of the Association for Computational Linguistics (Volume 1: Long Papers)",
    series = "ACL '25"
}

@inproceedings{fuzz4all,
    title = {{Fuzz4All}: Universal Fuzzing with Large Language Models},
    author = {Xia, Chunqiu Steven and Paltenghi, Matteo and Tian, Jia Le and Pradel, Michael and Zhang, Lingming},
    booktitle = {Proceedings of the 2024 46th International Conference on Software Engineering},
    year = {2024},
    series = {ICSE '24}
}

@inproceedings{metamut,
    title = {The Mutators Reloaded: Fuzzing Compilers with Large Language Model Generated Mutation Operators},
    author = {Ou, Xianfei and Li, Cong and Jiang, Yanyan and Xu, Chang},
    year = {2025},
    booktitle = {Proceedings of the 2024 29th ACM International Conference on Architectural Support for Programming Languages and Operating Systems, Volume 4},
    series = {ASPLOS '24}
}

@inproceedings{missedsizeopt,
    title = {Finding Missed Code Size Optimizations in Compilers using Large Language Models},
    author = {Italiano, Davide and Cummins, Chris},
    year = {2025},
    booktitle = {Proceedings of the 2025 34th ACM SIGPLAN International Conference on Compiler Construction},
    series = {CC '25}
}

@inproceedings{compilergym,
    title = {{CompilerGym}: Robust, Performant Compiler Optimization Environments for {AI} Research},
    author = {Cummins, Chris and Wasti, Bram and Guo, Jiadong and Cui, Brandon and Ansel, Jason and Gomez, Sahir and Jain, Somya and Liu, Jia and Teytaud, Olivier and Steiner, Benoit and Tian, Yuandong and Leather, Hugh},
    year = {2022},
    booktitle = {Proceedings of the 2022 20th IEEE/ACM International Symposium on Code Generation and Optimization},
    series = {CGO '22}
}

@inproceedings{llmcompiler,
    title = {{LLM Compiler}: Foundation Language Models for Compiler Optimization},
    author = {Cummins, Chris and Seeker, Volker and Grubisic, Dejan and Roziere, Baptiste and Gehring, Jonas and Synnaeve, Gabriel and Leather, Hugh},
    year = {2025},
    booktitle = {Proceedings of the 2025 34th ACM SIGPLAN International Conference on Compiler Construction},
    series = {CC '25}
}

@inproceedings{rl4real,
    title = {{RL4ReAl}: Reinforcement Learning for Register Allocation},
    author = {VenkataKeerthy, S. and Jain, Siddharth and Kundu, Anilava and Aggarwal, Rohit and Cohen, Albert and Upadrasta, Ramakrishna},
    year = {2023},
    booktitle = {Proceedings of the 2023 32nd ACM SIGPLAN International Conference on Compiler Construction},
    series = {CC '23}
}

@inproceedings{llmvec,
    title = {{LLM-Vectorizer}: {LLM}-Based Verified Loop Vectorizer},
    author = {Taneja, Jubi and Laird, Avery and Yan, Cong and Musuvathi, Madan and Lahiri, Shuvendu K.},
    year = {2025},
    booktitle = {Proceedings of the 2025 23rd ACM/IEEE International Symposium on Code Generation and Optimization},
    series = {CGO '25}
}

@inproceedings{phraseorderjikes,
    title = {Mitigating the Compiler Optimization Phase-Ordering Problem using Machine Learning},
    author = {Kulkarni, Sameer and Cavazos, John},
    year = {2012},
    booktitle = {Proceedings of the 2012 ACM International Conference on Object Oriented Programming Systems Languages and Applications},
    series = {OOPSLA '12}
}

@inproceedings{llmrlcodeperf,
    title={Improving Assembly Code Performance with Large Language Models via Reinforcement Learning},
    author={Anjiang Wei and Tarun Suresh and Huanmi Tan and Yinglun Xu and Gagandeep Singh and Ke Wang and Alex Aiken},
    year={2025},
    booktitle={Proceedings of the NeurIPS 2025 4th Workshop on Deep Learning for Code},
}

@inproceedings{rgym,
    title={Rethinking Kernel Program Repair: Benchmarking and Enhancing {LLM}s with {RGym}},
    author={Kareem Shehada and Yifan Wu and Wyatt D. Feng and Adithya Iyer and Gryphon Kumfert and Yangruibo Ding and Zhiyun Qian},
    year={2025},
    booktitle={Proceedings of the NeurIPS 2025 Workshop on Evaluating the Evolving LLM Lifecycle: Benchmarks, Emergent Abilities, and Scaling},
}

@inproceedings{kgym,
    title={{kGym}: A Platform and Dataset to Benchmark Large Language Models on {Linux} Kernel Crash Resolution},
    author={Alex Mathai and Chenxi Huang and Petros Maniatis and Aleksandr Nogikh and Franjo Ivancic and Junfeng Yang and Baishakhi Ray},
    year={2024},
    booktitle={Proceedings of the 2024 38th Conference on Neural Information Processing Systems Datasets and Benchmarks Track},
    series={NeurIPS '24}
}

@inproceedings{kernelbench,
    title={{KernelBench}: Can {LLM}s Write Efficient {GPU} Kernels?},
    author={Anne Ouyang and Simon Guo and Simran Arora and Alex L Zhang and William Hu and Christopher Re and Azalia Mirhoseini},
    year={2025},
    booktitle={Proceedings of the 2025 42nd International Conference on Machine Learning},
    series={ICML '25}
}

@inproceedings{react,
    title={{ReAct}: Synergizing Reasoning and Acting in Language Models},
    author={Shunyu Yao and Jeffrey Zhao and Dian Yu and Nan Du and Izhak Shafran and Karthik R Narasimhan and Yuan Cao},
    booktitle={Proceedings of the 2023 11th International Conference on Learning Representations},
    year={2023},
    series={ICLR '23}
}

@inproceedings{benchmarkprob,
    title={Benchmark Probing: Investigating Data Leakage in Large Language Models},
    author={Chunyuan Deng and Yilun Zhao and Xiangru Tang and Mark Gerstein and Arman Cohan},
    year={2024},
    booktitle={Proceedings of the NeurIPS 2023 Workshop on Backdoors in Deep Learning - The Good, the Bad, and the Ugly},
}

@inproceedings{locagent,
    title = "{LocAgent}: Graph-Guided {LLM} Agents for Code Localization",
    author = "Chen, Zhaoling  and
    Tang, Robert  and
    Deng, Gangda  and
    Wu, Fang  and
    Wu, Jialong  and
    Jiang, Zhiwei  and
    Prasanna, Viktor  and
    Cohan, Arman  and
    Wang, Xingyao",
    year = "2025",
    booktitle = "Proceedings of the 63rd Annual Meeting of the Association for Computational Linguistics (Volume 1: Long Papers)",
    series = "ACL '25"
}

@inproceedings{understandingcompilers,
    title = {Toward understanding compiler bugs in GCC and LLVM},
    author = {Sun, Chengnian and Le, Vu and Zhang, Qirun and Su, Zhendong},
    year = {2016},
    booktitle = {Proceedings of the 2016 25th International Symposium on Software Testing and Analysis},
    series = {ISSTA '16}
}

@article{yarpgen1,
    title = {Random testing for {C} and {C++} compilers with {YARPGen}},
    author = {Livinskii, Vsevolod and Babokin, Dmitry and Regehr, John},
    year = {2020},
    number = {OOPSLA},
    journal = {Proc. ACM Program. Lang.}
}

@article{yarpgen2,
    title = {Fuzzing Loop Optimizations in Compilers for {C++} and Data-Parallel Languages},
    author = {Livinskii, Vsevolod and Babokin, Dmitry and Regehr, John},
    year = {2023},
    number = {PLDI},
    journal = {Proc. ACM Program. Lang.}
}

@article{agentless,
    title = {Demystifying {LLM}-Based Software Engineering Agents},
    author = {Xia, Chunqiu Steven and Deng, Yinlin and Dunn, Soren and Zhang, Lingming},
    year = {2025},
    number = {FSE},
    journal = {Proc. ACM Softw. Eng.}
}

@article{creal,
    title = {Boosting Compiler Testing by Injecting Real-World Code},
    author = {Li, Shaohua and Theodoridis, Theodoros and Su, Zhendong},
    year = {2024},
    number = {PLDI},
    journal = {Proc. ACM Program. Lang.},
}

@article{whitefox,
    title = {{WhiteFox}: White-Box Compiler Fuzzing Empowered by Large Language Models},
    author = {Yang, Chenyuan and Deng, Yinlin and Lu, Runyu and Yao, Jiayi and Liu, Jiawei and Jabbarvand, Reyhaneh and Zhang, Lingming},
    year = {2024},
    number = {OOPSLA2},
    journal = {Proc. ACM Program. Lang.},
}

@article{legofuzz,
    title = {Interleaving Large Language Models for Compiler Testing},
    author = {Ni, Yunbo and Li, Shaohua},
    year = {2025},
    number = {OOPSLA2},
    journal = {Proc. ACM Program. Lang.},
}

@article{llm4vv,
    title = {{LLM4VV}: Developing {LLM}-Driven Testsuite for Compiler Validation},
    author = {Munley, Christian and Jarmusch, Aaron and Chandrasekaran, Sunita},
    year = {2024},
    journal = {Future Gener. Comput. Syst.},
}

@article{autotunesurvey,
    title = {A Survey on Compiler Autotuning using Machine Learning},
    author = {Ashouri, Amir H. and Killian, William and Cavazos, John and Palermo, Gianluca and Silvano, Cristina},
    year = {2018},
    journal = {ACM Comput. Surv.},
}

@article{reify,
    author = {Chopra, Kavya and Li, Cong and Sotiropoulos, Thodoris and Su, Zhendong},
    title = {Semantic Reification: A New Paradigm for Random Program Generation},
    year = {2026},
    number = {PLDI},
    journal = {Proc. ACM Program. Lang.},
}

@inproceedings{missedpeepholeopt,
    author = {Xu, Zhenyang and Xu, Hongxu and Tian, Yongqiang and Zhou, Xintong and Sun, Chengnian},
    title = {{LPO}: Discovering Missed Peephole Optimizations with Large Language Models},
    year = {2026},
    booktitle = {Proceedings of the 2026 31st ACM International Conference on Architectural Support for Programming Languages and Operating Systems},
    series = {ASPLOS '26}
}

@misc{mlgo,
    title={{MLGO}: a Machine Learning Guided Compiler Optimizations Framework}, 
    author={Mircea Trofin and Yundi Qian and Eugene Brevdo and Zinan Lin and Krzysztof Choromanski and David Li},
    year={2021},
    eprint={2101.04808},
    archivePrefix={arXiv},
    primaryClass={cs.PL},
    url={https://arxiv.org/abs/2101.04808}, 
}

@misc{supercode,
    title={{SuperCoder}: Assembly Program Superoptimization with Large Language Models}, 
    author={Anjiang Wei and Tarun Suresh and Huanmi Tan and Yinglun Xu and Gagandeep Singh and Ke Wang and Alex Aiken},
    year={2025},
    eprint={2505.11480},
    archivePrefix={arXiv},
    primaryClass={cs.CL},
    url={https://arxiv.org/abs/2505.11480}, 
}

@misc{crashfixer,
    title={{CrashFixer}: A Crash Resolution Agent for the {Linux} Kernel}, 
    author={Alex Mathai and Chenxi Huang and Suwei Ma and Jihwan Kim and Hailie Mitchell and Aleksandr Nogikh and Petros Maniatis and Franjo Ivančić and Junfeng Yang and Baishakhi Ray},
    year={2025},
    eprint={2504.20412},
    archivePrefix={arXiv},
    primaryClass={cs.SE},
    url={https://arxiv.org/abs/2504.20412}, 
}

@online{swebenchverified,
    author = {OpenAI},
    year   = {2024},
    title  = {Introducing {SWE}-bench {V}erified},
    url    = {https://openai.com/index/introducing-swe-bench-verified/},
    note = {Accessed: Mar 17th 2026}
}

@online{geminicli,
    author = {Gemini},
    year   = {2025},
    title  = {Gemini {CLI}},
    url    = {https://github.com/google-gemini/gemini-cli},
    note = {Accessed: Mar 17th 2026}
}

@online{codex,
    author = {OpenAI},
    year   = {2025},
    title  = {Codex},
    url    = {https://github.com/openai/codex},
    note = {Accessed: Mar 17th 2026}
}

@online{minisweagent,
    author = {{SWE}-agent},
    year   = {2024},
    title  = {mini-swe-agent},
    url    = {https://github.com/SWE-agent/mini-swe-agent},
    note = {Accessed: Mar 17th 2026}
}

@online{llvmlangref,
    author = {{LLVM}},
    year   = {2026},
    title  = {{LLVM} Language Reference Manual},
    url    = {https://llvm.org/docs/LangRef.html},
    note = {Accessed: Mar 17th 2026}
}

@online{llvmpasses,
    author = {{LLVM}},
    year   = {2026},
    title  = {{LLVM}'s Analysis and Transform Passes},
    url    = {https://llvm.org/docs/Passes.html},
    note = {Accessed: Mar 17th 2026}
}

@online{llvmaiguide,
    author = {{LLVM}},
    year   = {2026},
    title  = {{LLVM} {AI} Tool Use Policy},
    url    = {https://llvm.org/docs/AIToolPolicy.html},
    note = {Accessed: Mar 17th 2026}
}

@online{swebenchleaderboard,
    author = {{SWE}-bench},
    year   = {2024},
    title  = {SWE-bench Leaderboards},
    url    = {https://www.swebench.com/index.html},
    note = {Accessed: Mar 17th 2026}
}

@online{openaigiveupswe,
    author = {{O}pen{AI}},
    year   = {2026},
    title  = {Why SWE-bench Verified No Longer Measures Frontier Coding Capabilities},
    url    = {https://openai.com/index/why-we-no-longer-evaluate-swe-bench-verified/},
    note = {Accessed: Mar 17th 2026}
}

@online{dsv32issue1,
    author = {v{LLM}},
    year   = {2025},
    title  = {[Bug]: Deepseek V3.2 tool\_calls Failure},
    url    = {https://github.com/vllm-project/vllm/issues/26897},
    note = {Accessed: Mar 17th 2026}
}

@online{dsv32issue2,
    author = {v{LLM}},
    year   = {2025},
    title  = {[Bug]: Frequent Tool Call Parsing Failures with DeepSeek-V3.2},
    url    = {https://github.com/vllm-project/vllm/issues/36654},
    note = {Accessed: Mar 17th 2026}
}

@online{dsv32issue3,
    author = {{SGL}ang},
    year   = {2025},
    title  = {[Bug] DeepSeek-V3.2 Occasional Malformed Tool Call Output},
    url    = {https://github.com/sgl-project/sglang/issues/14695},
    note = {Accessed: Mar 17th 2026}
}

@online{dsv32issue4,
    author = {{SGL}ang},
    year   = {2026},
    title  = {[Bug] DeepSeek-V3.2 Places Tool Call Content in "content" Field Instead of "tool\_calls" When Invoking Tools, and finish\_reason is "stop" Rather Than "tool\_calls"},
    url    = {https://github.com/sgl-project/sglang/issues/17561},
    note = {Accessed: Mar 17th 2026}
}

@online{llvmreviewprinciple,
    author = {{LLVM}},
    year   = {2026},
    title  = {LLVM Code-Review Policy and Practices},
    url    = {https://llvm.org/docs/CodeReview.html},
    note = {Accessed: Mar 17th 2026}
}

@techreport{contextrot,
    author = {Hong, Kelly and Troynikov, Anton and Huber, Jeff},
    title = {Context Rot: How Increasing Input Tokens Impacts {LLM} Performance},
    institution = {Chroma},
    year = {2025}
}
